\documentclass[journal]{IEEEtran}

\ifCLASSINFOpdf
\else
\fi
\usepackage{pdfsync}

\usepackage{multicol,color}
\usepackage{ragged2e}
\usepackage{amssymb}
\usepackage{amsmath}
\usepackage{tabularx}
\usepackage{booktabs}
\usepackage{ntheorem}
\usepackage{multicol}
\usepackage{makecell}
\usepackage{graphicx}

\newtheorem{theorem}{\textbf{Theorem}}

\newtheorem{lemma}{\textbf{Lemma}}
\newtheorem{remark}{\textbf{Remark}}
\newtheorem{proposition}{\textbf{Proposition}}

\newtheorem{definition}{\textbf{Definition}}
\usepackage{graphics}
\usepackage{algorithm}
\usepackage{algorithmicx}
\usepackage{algpseudocode}
\usepackage{subfigure}

\usepackage[]{hyperref}

\hypersetup{
	colorlinks=true,
	linkcolor=blue,
	filecolor=blue,      
	urlcolor=blue,
    citecolor=blue,
}

\usepackage{array}
\usepackage{url}

\usepackage{graphicx}

\makeatletter

\renewcommand{\maketag@@@}[1]{\hbox{\m@th\normalsize\normalfont#1}}%

\makeatother
\newtheorem{assumption}{\textbf{Assumption}}

\makeatletter

\begin{document}
%
\title{Distributed Stackelberg Equilibrium Seeking for Networked Multi-Leader Multi-Follower Games with A Clustered Information Structure
}
%
%

\author{
	Yue Chen and Peng Yi

\thanks{The paper was sponsored by 
	 the National Key Research and Development Program of China under No 2022YFA1004700.
	 \textit{(Corresponding author: Peng Yi.)}}	 
\thanks{The authors are with the Department of Control Science and Engineering,
	Tongji University, Shanghai, 201804, China, and also with  Shanghai Institute of Intelligent Science and Technology, Tongji University, Shanghai 200092, China. (email: chenyue\_j@tongji.edu.cn, yipeng@tongji.edu.cn).  }
}

\markboth{ }
{Shell \MakeLowercase{\textit{et al.}}: Bare Demo of IEEEtran.cls for IEEE Journals}

\maketitle

\begin{abstract}

The Stackelberg game depicts a leader-follower relationship wherein decisions are made sequentially, and the Stackelberg equilibrium represents an expected optimal solution when the leader can anticipate the 
rational response of the follower.
Motivated by  control of network systems with two levels of decision-making hierarchy, such as the management of energy  
networks and power coordination at  cellular networks,  
a networked multi-leaders and multi-followers Stackelberg game is proposed.
Due to the constraint of limited information interaction among players, a clustered information structure is assumed that each leader can only communicate with a portion of overall followers, namely its subordinated followers,  and also only with its local neighboring leaders. 
In this case, the leaders cannot fully anticipate the 
collective rational response of all followers with its local information.
To address Stackelberg equilibrium seeking under this partial information structure, we propose a distributed seeking algorithm based on implicit gradient estimation and network consensus mechanisms. 
We rigorously prove the convergence of the algorithm for both diminishing and constant step sizes under strict and strong monotonicity conditions, respectively.
Furthermore, the model and the algorithm can also incorporate linear equality and inequality constraints into the followers' optimization problems, with the approach of the interior point barrier function.
Finally, we present numerical simulations in  applications to corroborate our claims on the proposed framework.
\end{abstract}

\IEEEpeerreviewmaketitle

\section{Introduction}
\par 
Non-cooperative games serve as a prevalent mathematical framework for modeling systems that involve interacting strategic agents, where each agent competes independently to optimize its utility.
Among these, the Stackelberg game has garnered significant attention from researchers \cite{von2010market}. In the Stackelberg game, players are categorized as leaders and followers, with leaders taking the initiative decision and followers making their optimal decisions after observing leaders' moves.
This game model provides an effective framework for analyzing and solving decision problems when the decision-makers are in different hierarchies and with asymmetric information structures.
The Stackelberg game model has generated many variations, including
one-leader one-follower (OLOF), one-leader multi-follower (OLMF) \cite{fabiani2021local}, multi-leader one-follower (MLOF) \cite{xie2023multiplayer}, and multi-leader multi-follower (MLMF) game models \cite{hu2015multi}, 
	where the MLMF game encompasses all other variations and fits well with hierarchical decision-making in various networks, 
	such as the internet of things \cite{ding2020incentive}, cellular network systems \cite{zhang2016multi}, and smart grids \cite{chen2017stackelberg}.
\par 
The Stackelberg equilibrium (SE) is commonly used to describe a stationary state that is achieved between rational leaders and followers in Stackelberg games, in which the leaders and followers have no incentive to unilaterally deviate from their respective decisions \cite{fabiani2021local}.  
In SE seeking settings, the full information  in the context means that the leader can anticipate the rational response of the followers, thus achieving a {\it first-mover-advantage.} 
\par Both the Mathematical Program with Equilibrium Constraints (MPEC) approach and the bilevel optimization approach have been used in  the SE seeking.
In the MPEC approach for OLMF games,  the followers' optimization problem is first transformed into a variational inequality (VI) with KKT analysis,
and then an equilibrium problem with equilibrium constraints is formulated and solved \cite{leyffer2003mathematical,naebi2020epec}. 
In the bilevel optimization approach for OLOF games, 
the optimal response function of the follower is incorporated 
into the leader's optimization problem.
When the leader has full information of the follower, the closed-form expression of the follower's best response function can be used by the leader for SE computation, such as in \cite{von2010leadership,kebriaei2017discrete}. 
On the other hand,  when the leader does not have   full information of the follower, 
the leader can obtain an estimate of the gradient of the best response function by exchanging information with the follower, with the help of implicit differentiation gradient approximating method in bilevel optimization \cite{ji2021bilevel}.
%
%
%
However, the above computational approaches are usually utilized to handle the 
Stackelberg games with only one leader.
Moreover, for MLMF games, {\it  they do not consider the partial information constraint that each player can only have information exchange with a portion of overall players, which is common in networks.
%
 In fact, as the number of players increases and the network scale expands, it would be impossible for each leader to communicate with all followers in networked systems. }
\par Hence, we impose a clustered information structure for the MLMF game of networked players to capture the limited information interaction between leaders and followers. 
Each leader has neighboring leaders and directly subordinated followers, while each follower only communicates with their superior leader. 
Such a communication paradigm operates by having leaders serve as coordinators responsible for receiving the information of their subordinate followers and are responsible for local inter-leader communications.
The same information structure is also employed in the applications of 
	many multi-agent systems, such as 
	vehicle networking \cite{niyato2010optimal}, wireless sensor networks \cite{rawat2021clustering}, etc.
	Furthermore, wireless communication modalities and networking, such as cellular networks (4G, 5G) and Wireless Fidelity (Wi-Fi), are particularly suitable for implementation of the clustered information scheme \cite{zhang2011game}.
\par  Then the challenges arise for seeking the SE with such a clustered information structure. Firstly, each leader cannot directly get the rational response of all the followers, since it only communicates with a portion of followers, namely its subordinated followers. Secondly, the leaders also need to compete in a noncooperative game with only local information exchange.
In this work, we propose an innovative approach that combines implicit gradient estimation and network consensus techniques to achieve a SE for a networked MLMF game.
\par Our main contributions are listed as follows
\begin{enumerate}
	\item 
    We propose a distributed SE seeking algorithm for a networked MLMF Stackelberg game with a cluster information structure. A novel approach for estimating the followers' global best response function
    is designed by combining a  gradient approximation approach and a network consensus method. 
    Note that our method is applicable to networked multi-entity systems,
    and is able to implemented in a distributed manner. 
    In contrast, most existing research primarily focuses on  
    single-leader (follower) scenarios, and the algorithms employed are typically assumed with full information. 
	\item We perform a convergence analysis of the algorithm. Under a certain condition on the 
number of inner loops, we show that the algorithm can converge to the SE with a vanishing step size and a constant step size with proper pseudo-gradient assumptions. 
	We also utilize interior point barrier function approach to solve the case when the followers' optimization problems have linear equality and inequality constraints.	
\end{enumerate}
\par The paper is organized as follows. Section \ref{section2} formulates networked multi-leader multi-follower games with a clustered information structure. Sections \ref{section3} and \ref{section4} introduce a SE seeking algorithm and its convergence analysis, respectively. Section \ref{section5} presents the numerical simulations, and Section \ref{section6} concludes the paper.
\subsection{Notation and Preliminaries} \label{section1}
Let $m$ be a positive integer, and denote the  $ m $-dimensional (non-negative) Euclidean space by $\mathbb{R}^m(\mathbb{R}^m_+) $.
In this paper, all vectors are viewed as column vectors.
Let $ \boldsymbol{1}_m $ and $ \boldsymbol{0}_m $ denote the $ m $-dimensional vectors with all elements being $ 1 $ and $ 0 $.
The $ m \times m $ identity matrix is denoted by $ \boldsymbol{I}_m $.
Denote $ x^{\top} $ and $ A^{\top} $ as the transpose of a vector $ x $ and a matrix $ A $.
A set composed of positive integers ranging from $1$ to $m$ is denoted by $\{1,2,\dots,m\} $.
If this set is $\mathcal{I}$, then 
the notation $ \text{col}((x^j)_{j\in \mathcal{I}})$ is the same as  $ [(x^1)^{\top}, \dots, (x^m)^{\top} ]^{\top} $.
For the matrices $ A^1, \dots, A^m  $,  $ \text{diag}((A^j)_{j\in \mathcal{I}}) $ denotes the block diagonal matrix with $ A^1, \dots, A^m  $ in the block diagonal positions.
Denote by $ [A]_{rs}$ the item of $ r $-th row $ s $-th column of matrix $ A$ and $ [x]_s $ the $ s $-th item of vector $ x$.
Let $ \| \cdot \| $, $ \otimes $, $ \left\langle x, y \right\rangle $ be the standard Euclidean norm (or induced matrix norm),
the Kronecker product,
and the standard inner product of $ x,y\in \mathbb{R}^m $ respectively.
Denote by $ \{\beta_k\}^m $ a sequence $ \beta_k $ ranging from $ k=0 $ to $ m $.
For any $j\in \{1,2,\dots, m\}$,  a function $ f(x^1, . . . , x^m,y) $, denote by $ \nabla_{x^j} f $,  $ \nabla_y\nabla_{x^j} f$, and $ \nabla^2_y f$ the gradient of $ f $ w.r.t. $ x^j $, the transpose of Jacobian matrix  w.r.t. $ (x^j, y) $, and the Hessian matrix w.r.t. $y$, respectively.
\section{PROBLEM FORMULATION} \label{section2}
\subsection{ MLMF Stackelberg Games}\label{formulation}
\par Consider an MLMF Stackelberg game with $m+N$ players, where the set of all players is denoted as $ \mathcal{I}:=\{ 1,2,\dots, m+N\}$, with the set of leaders as $ \mathcal{I}_L :=\{1,2,\dots, m\} $ and the set of followers as $\mathcal{I}_F:=\{m+1,m+2,\dots, m+N\}$. It holds that $\mathcal{I}= \mathcal{I}_L \cup \mathcal{I}_F$.
For any	 $i\in \mathcal{I}_F$,  follower $i$ has a decision variable $y^i \in \Omega_i \subseteq \mathbb{R}^{p_i}$, and aims to optimize the following problem when leaders' collective decision variable $x$ is given:
\small
\begin{equation} \label{followergame}
	\left\lbrace 
	\begin{aligned}
		\underset{ y^i } {\operatorname{min}}  \quad& 
		s^i (y^i, x ) \\ 
		s.t.  \quad 
		&  y^i \in \Omega_i .\\
	\end{aligned}
	\right. 
\end{equation}
\normalsize 
\par Meanwhile, 
for any $ j\in \mathcal{I}_L $, leader $j$
has its decision variable 
$ x^j \in  \Omega_j \subseteq  \mathbb{R}^{q_j} $ and aims to solve the form of optimal  problem as follows:
\small
\begin{equation} \label{leaderproblem}
	\left\lbrace 
	\begin{aligned}
		\underset{ x^j} {\operatorname{min}}  \quad 
		 &\theta^j (x^j, x^{-j},  y^*(x) ) \\ 
		s.t. \quad  &x^j \in \Omega_j,
	\end{aligned}
	\right. 
\end{equation}
\normalsize
where $y^*(x):=\text{col}(y^{1,*}(x),\dots,y^{N,*}(x)) $ is a collective vector of all followers, 
and $ i $-th item  $ y^{i,*}(x) $ is the optimal solution of follower $ i $, i.e., $y^{i,*}(x) := \text{argmin}_{y^i\in \Omega_i}\,\, s^i(y^i,x)$.
It is worth mentioning that $y^{i,*}(x) $ is a function w.r.t. $x$.
Moreover, $x^{-j} := \text{col}((x^l)_{l\in \mathcal{I}_L,l\neq j})$, $ x:=\text{col}((x^j)_{j\in \mathcal{I}_L}) \in \mathbb{R}^{q:=\sum_{j=1}^{m}q_j} \in \Omega$, and $ \Omega=\Omega_1\times\dots\times\Omega_m $.
\par Correspondingly, an MLMF Stackelberg game $ \left\langle  \mathcal{I}_L, \mathcal{I}_F,
 \Omega_j, \Omega_i, \theta^j, s^i \right\rangle     $ is defined as follows:
\begin{equation} \label{gamemodel}
	\begin{cases}
		& \text{Players:}    \quad  \text {Leaders set:} \, \mathcal{I}_L,\,\, \text{Followers set:}\, \mathcal{I}_F,       \\
		& \text{Strategies:}  
		 \begin{cases}
			&\text{Leader}\,   j:  \,\,\,  x^j \in  \Omega_j ,  \\
			&\text{Follower}\, i:  \,\,  y^i  \in \Omega_i ,  \\
		\end{cases}\\
		& \text{Cost:} 
		\begin{cases}
			&\text{Leader}\,   j:  \,\,  \theta^j (x^j,x^{-j}, y^*(x)) \,\text{defined}\,\text{in} \, (\ref{leaderproblem}), \\
			&\text{Follower}\, i:  \,  s^i (y^i,x) \,\text{defined} \,\text{in} \, (\ref{followergame}).  \\
		\end{cases}
	\end{cases}
\end{equation} 
\par 
Specifically, we focus on a case where 
$ \Omega_i=\mathbb{R}^{p_i}, \forall i \in \mathcal{I}_F $, while we also discuss the feasible domain consisting of the intersection of inequality and equality constraints in Part \ref{subsectionofgeneralized}.
In the following, we impose assumptions about cost functions of all players, including leaders and followers.
\begin{assumption} \label{assumptionoffunction}
	For each leader $ j\in \mathcal{I}_L $, the cost function 
	$ \theta^j(x^j, x^{-j}, y^*(x)) $ is convex w.r.t. $ x^j $ on $ \Omega_j $, where $ \Omega_j $ is convex and compact,
	and $ \theta^j(x^j, x^{-j}, y) $ is continuously differentiable w.r.t. $ (x^j, y) $ on $ \Omega_j \times \mathbb{R}^p $.
	Furthermore, for each follower $ i\in \mathcal{I}_F $, the cost function 
	$ s^i(y^i, x) $ is $ \mu $-strongly convex  w.r.t. $ y^i $ on  $ \mathbb{R}^{p_i} $, and twice continuously differentiable w.r.t. $ (y^i, x) $ on $  \mathbb{R}^{p_i} \times \Omega $. 
\end{assumption}
\par By the strongly convexity property of follower $i \in \mathcal{I}_F$ in the above assumption, the existence and uniqueness of $y^{i,*}(x)$ are readily derived. Furthermore, we introduce
 some mild continuous assumptions on players' cost functions, which is common in bilevel optimization , such as \cite{ghadimi2018approximation}.
\begin{assumption} \label{assumption_lipschitz}
	For every $ j\in \mathcal{I}_L $, $ \theta^j(x^j,x^{-j},y) $ and $ \nabla \theta^j(x^j,x^{-j},y)  $ are 
	$ \ell_{\theta,0} $ and $ \ell_{\theta,1} $ Lipschitz continuous w.r.t. $ (x^j, y) $ on $ \Omega_j \times \mathbb{R}^p $ with fixed  $ x^{-j} $.
	For every $ i\in \mathcal{I}_F $, 
	$ s^i  (y^i,x) $,  
	$ \nabla s^i  (y^i,x)$, 
	and $ \nabla^2 s^i  (y^i,x) $
	are $ \ell_{s,0} $, $ \ell_{s,1} $ and  $ \ell_{s,2} $ Lipschitz continuous  
	w.r.t. $ (y^i,x) $ on $ \mathbb{R}^{p_i} \times \Omega $.
	Moreover, we define $ \ell = \text{max}\{ \ell_{\theta,1}, \ell_{s,1}   \} $, and $ \kappa:=\frac{\ell}{\mu} $.
\end{assumption}
\par 
Assumption \ref{assumptionoffunction} and \ref{assumption_lipschitz} address the properties of the players' cost functions in the model, establishing theoretical guarantees for the existence analysis of SE.
In an MLMF Stackelberg game, leaders compete with each other and generate optimal decisions while being aware of followers' responses.
Once the leaders have made their decisions, the followers observe them and generate their optimal decisions. 
Herein, we introduce the definition of an MLMF Stackelberg equilibrium  as followers.
\begin{definition}(Stackelberg equilibrium)
	For any $ j \in \mathcal{I}_L $, $ x^{j,\diamond} $ is a Stackelberg strategy for leader $ j $, if 
	\small
	\begin{flalign}
		\theta^j (x^{j,\diamond},x^{-j,\diamond},y^*(x^{j,\diamond},x^{-j,\diamond})) \leq \theta^j (x^j, x^{-j,\diamond}, y^*(x^j,x^{-j,\diamond})),  
		\notag
	\end{flalign}
    \normalsize
	where $ x^{\diamond}=(\text{col}(x^{j,\diamond})_{j\in \mathcal{I}_L} )$ is a Nash equilibrium for leaders, 
	and for any follower $ i \in \mathcal{I}_F $ with given $ x^{\diamond} $,
	\small
	\begin{flalign}
 s^i (y^{i,\diamond},x^{\diamond}) \leq s^i (y^i,x^{\diamond}), \notag
	\end{flalign}
    \normalsize
which indicates that $ y^{\diamond} $ is an optimal response of followers. Above all, $ (x^{\diamond}, y^{\diamond}) $ is a Stackelberg equilibrium of the proposed game. 
\end{definition} 
\par Note that by the definition of the best response function
 of follower $i$ w.r.t. $x$ and the SE, it can be derived that $y^{i,\diamond} =  y^{i,*}(x^\diamond) $, and $ y^{\diamond} =  y^{*}(x^\diamond) $.
 \subsection{Networked MLMF Stackelberg Games } \label{topology}
\par In this subsection,
we introduce a networked MLMF game
$ \langle  \mathcal{I}_L, \mathcal{I}_F,
\Omega_j, \Omega_i, \theta^j, s^i, \tilde{\mathcal{G}}  \rangle     $, where $ \tilde{\mathcal{G}}$ denotes the clustered information structure of a networked system.
\par
\begin{figure}[htbp]  
		\centering
		\includegraphics[width=3.3in]{ 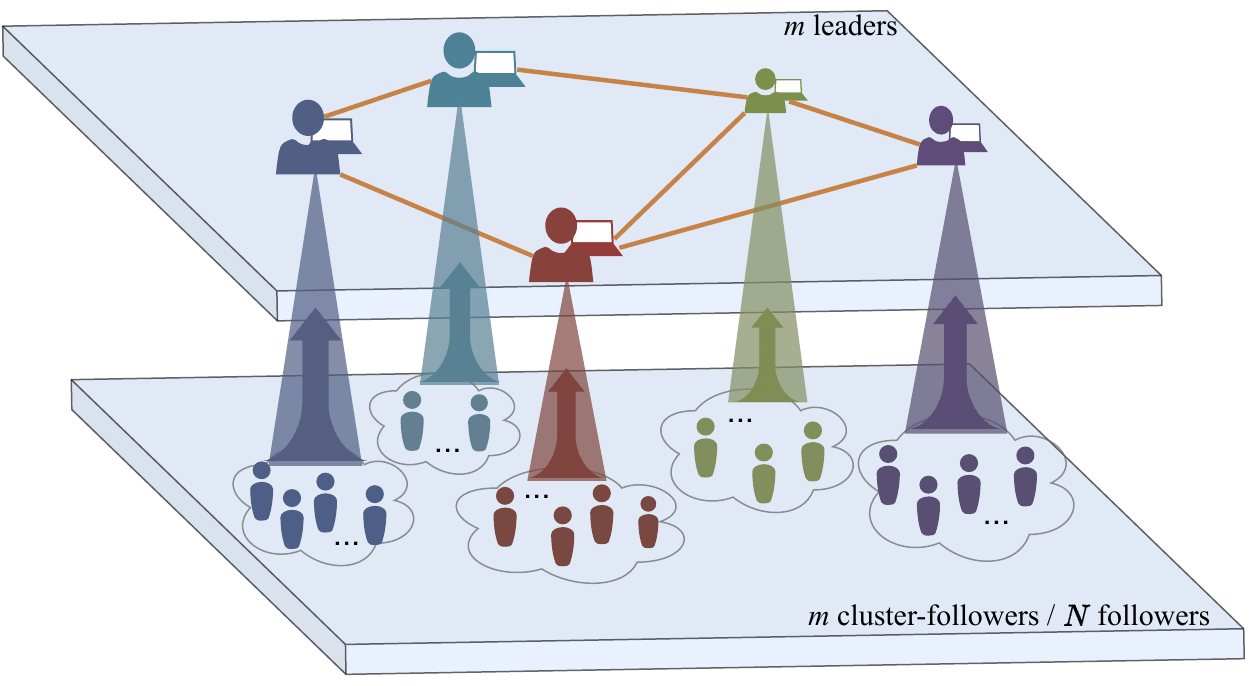}\\
		\caption{Clustered Information Structure. }\label{clustered}
	\end{figure}

 The clustered information structure $\tilde{\mathcal{G}}$  is  illustrated in Fig. \ref{clustered}, where the communication network among leaders follows an undirected graph topology $\mathcal{G}$, and  followers communicates with each leader forming a cluster.
To better clarify the dependency relationship between leaders and followers in Stackelberg games, we adopt the interference graph setting discussed in \cite{yi2019operator}. This setting implies that all players' decision variables can be obtained through observation. 
\par The clustered information structure  $\tilde{\mathcal{G}}$
allows  each leader only to have transmitted information of followers belonging to its cluster, which imposes challenges 
seeking SE. 
A network consensus approach is required to 
help the leaders to estimate the unknown  global information.
	\par 
\par We denote by 
	$ \mathcal{G}:=(\mathcal{I}_L,\mathcal{E})$  an undirected graph of leaders, where $ \mathcal{I}_L $ is the set of agents, $\mathcal{E}\subset \mathcal{I}_L\times \mathcal{I}_L$ is the edge set,
	and $ W:=[w_{jg}]\in \mathbb{R}^{m\times m} $ is the adjacency matrix.
	If $ w_{jg} = w_{gj} >0 $, it means agents $j,g$ can mutually exchange information,
	and if  $ w_{jg }=0 $, it means that agents $ j,g $ are not able to exchange information directly.
	The set of neighbors of agent $ j $  is defined as $ \mathcal{N}_{jg} = \{g| (j, g) \in  \mathcal{E}\} $.
	We denote  $ d_{j} := \sum_{g=1}^{m} w_{jg} $,
	and $ \text{Deg} := \text{diag}((d_{j})_{j\in \mathcal{I}_L})\in\mathbb{R}^{m \times m} $.
	The weighted Laplacian of $ \mathcal{G} $ is defined as $ L := \text{Deg}-W $.
	For $ j,g\in \mathcal{I}_L $, if there exists a sequence of distinct nodes  $ j,j_1,\dots,j_p,g $ such that
	$ (j,j_1) \in \mathcal{I}_L\times \mathcal{I}_L $,
	$ (j_1,j_2) \in \mathcal{I}_L\times \mathcal{I}_L $,
	$ \dots $,
	$ (j_p,g) \in \mathcal{I}_L\times \mathcal{I}_L$,
	then we call $ (j,j_1, \dots, j_p,g)  $ the undirected path between $ j $ and $ g $.
	If there exists an undirected path between any $ j,g\in \mathcal{I}_L $, then $  \mathcal{G} $ is connected.
	We impose the following assumption on the communication network.
	\begin{assumption} \label{assumption_graph}
		Graph $ \mathcal{G} $ is undirected and connected.
		The adjacency matrix $ W $ of $ \mathcal{G} $ are double-stochastic 
		with positive diagonal elements, i.e.,
		$  W \boldsymbol{1}_m=\boldsymbol{1}_m $,
		$ \boldsymbol{1}^T_m W=\boldsymbol{1}_m^T $,
		$ w^{jj} >0 $, $ j\in \mathcal{I}_L $.
	\end{assumption}
	\par Assumption \ref{assumption_graph} is a typical setting for networked systems and 
	 it can be fulfilled by using the Metropolis-Hastings rule.
   \subsection{Examples:}
	 \par
\begin{figure}[htbp] 
	 	\centering
	 	\includegraphics[width=3.3in]{ 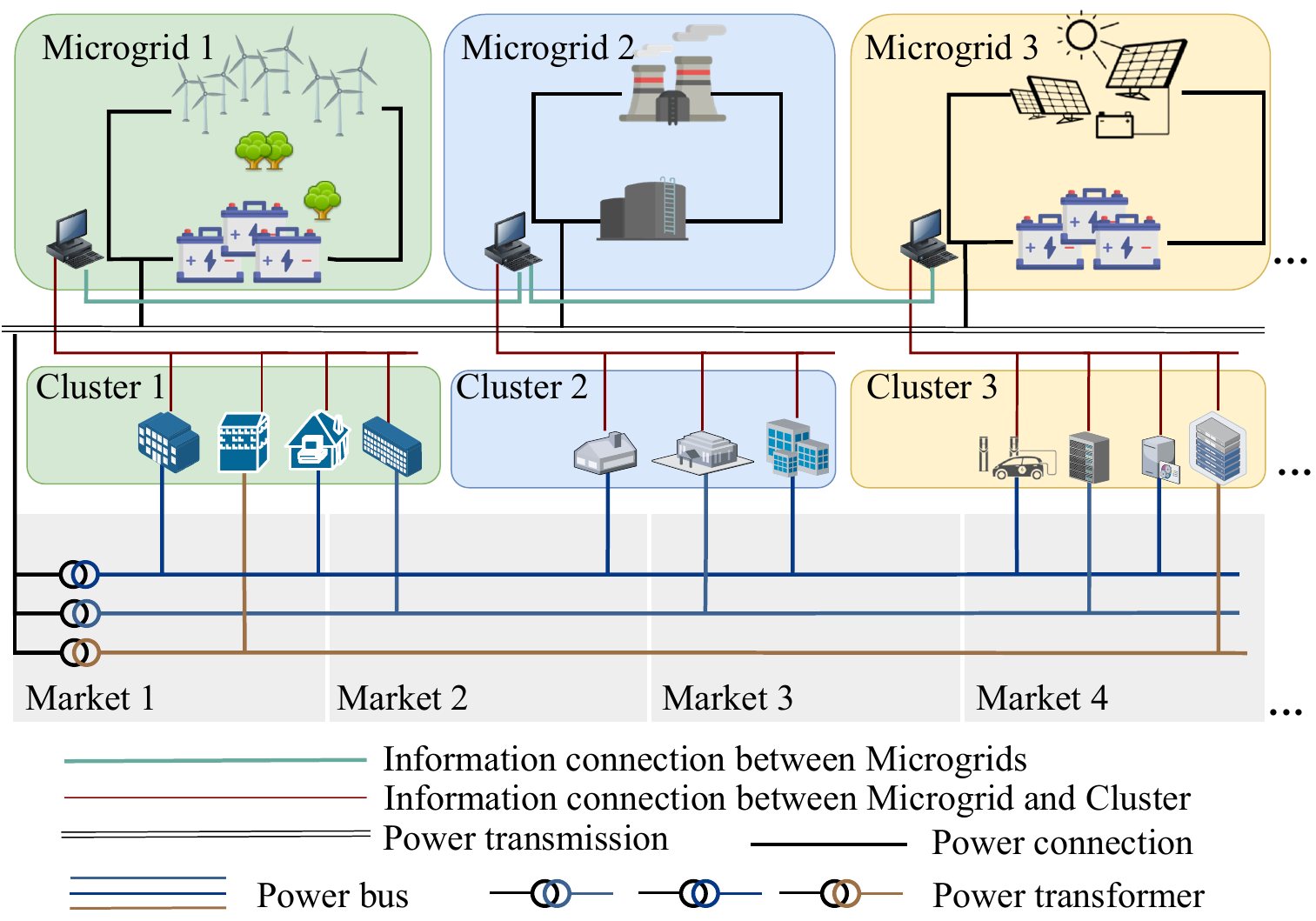}\\
	 	\centering
	 	\caption{Schematic of a microgrid management system.}\label{MG}
	 \end{figure}

 \textit{1) Networked Stackelberg-Cournot Equilibrium Problems:} The microgrid management system is shown in Fig.\ref{MG}. The figure illustrates a microgrid system with wind, solar, and chemical energy as the primary sources of power. 
	 The microgrids are physically connected to each other through power transmission buses, as depicted by the black double horizontal lines. 
	\par Considering the optimal response of demand-side users, the non-cooperative microgrids compete in the four demand-side markets, resulting in an equilibrium between the microgrids and all demand-side users.
	 Thus, the cost function of microgrid $ j\in \mathcal{I}_L $ is given by  
	 \small
	 \begin{equation} \label{microgrid}
	 	\begin{split}
	 		&\theta^j (x^j, x^{-j}, y) =  \frac{1}{2} (x^j) ^\top Q^j x^j +(d^j)^\top  x^j + u  \\
	 		&- (e-P\boldsymbol{A}x)^\top A^j x^j  + \bar{r}  \| \boldsymbol{A}x-\boldsymbol{B}y \|^2,    
	 	\end{split}
	 \end{equation}
	 \normalsize
	 where $ x^j \in \Omega_j $ is the power generation decision of microgrid $ j $, 
	 $ y $ is the optimal demand of users,
	 $ Q^j \in \mathbb{R}^{q_j \times q_j  } $, 
	 $ d^j \in \mathbb{R}^{q_j} $, $ e \in \mathbb{R}^v $, 
	 $ A^j \in \mathbb{R}^{v \times q^j} $,
	 $ \boldsymbol{A}:=[A^1, \dots, A^m] \in \mathbb{R}^{v \times q } $,
	 $ P\in \mathbb{R}^{v\times v} $,
	 $ u \in \mathbb{R} $, 
	 $ \bar{r}\in \mathbb{R} $,
	 $ \boldsymbol{B}:=[B^{1}, \dots, B^{N}] \in \mathbb{R}^{v \times p} $,
	 and $ B^{i} \in \mathbb{R}^{v \times p_i} $.
	 Likewise, the cost function of user $ i \in \mathcal{I}_F $ is given by
	 \small 
	 \begin{equation}
	 	\begin{split}
	 		&s^{i} ( y^{i} , x  ) =  \frac{1}{2} (  y^{i})^\top M^{i}   y^{i}  - (g^{i})^\top   y^{i} + (e-P\boldsymbol{A}x)^\top B^iy^{i},  \notag
	 	\end{split}
	 \end{equation}
	 \normalsize
	 where $ M^{i}  \in \mathbb{R}^{p_i \times p_i }$ 
	 and 
	 $g^{i} \in \mathbb{R}^{p_i}  $. 
	 The definition of these variables can be seen in \cite{yi2019operator, liang2019generalized}. Moreover,  
	 $ A^j  $ specifies the participation of microgrid $ j\in \mathcal{I}_L $ in the markets, 
	 and $ [A^j]_{rs} =1  $ implies that microgrid $ j $ provides $ [x^j]_s $ power to market $r$. 
	 Moreover, $ \boldsymbol{A}x = \sum_{j=1}^{m} A^j x^j $ indicates the total supply vector to all markets.
	 Note that any column of $ A^j $ only has one entry equal to $ 1 $, while the other entries are $ 0 $.
	 The same definitions apply to $ B^{i} $ and $ \boldsymbol{B} $, which specify  the demand of user $ i $ in cluster $ j $
	 and the total demand of all users in the markets. 
	 The last term of (\ref{microgrid}) regulates $\bar{r}$ to  make the total generation of microgrids is as close as possible to the total market demand.
	 \par
	As illustrated in Fig.\ref{MG}, a clustered information structure emerges for addressing the communication requirements for connecting all microgrids and each demand-side user, especially when considering the cost of constructing information lines for users in practical applications. The green lines in Fig.\ref{MG} establish connections between microgrids, while the red lines facilitate communication between clusters and demand users.
	 \par \textit{2) A Class of Heterogeneous Cellular Networked System:}
	 We discuss a heterogeneous cellular networked system with $ m $ operators and $ N $ users, as introduced in  \cite{zhang2016multi}, where each operator has unlicensed and licensed spectrum, and each user connects to any spectrum based on its needs. 
	 To discourage users from continuously seeking unlicensed spectrum resources, each operator $ j \in \mathcal{I}_L $ needs to set a corresponding penalty price $x^j $, while each user $i\in \mathcal{I}_F$ determines its optimal transmit power $ y^i $ upon receiving the price. 
	 \begin{figure}[htbp] 
	 	\centering
	 	\includegraphics[width=3.5in]{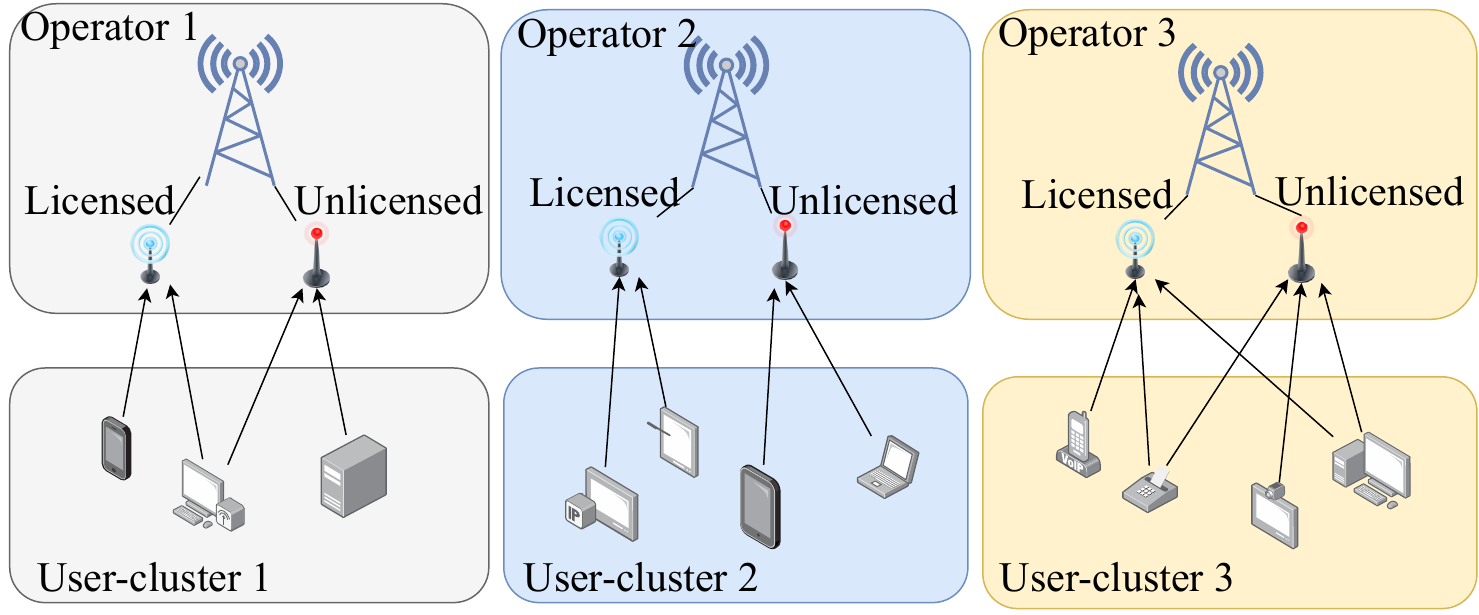}\\
	 	\caption{Schematic of a heterogeneous cellular networked system. }\label{WIFI}
	 \end{figure}
	 \par The spectrum efficiency function is defined as $ R(y^i):= \log(a^i+ (z^i)^\top y^i   )$, then
	 the cost function of user $i$  is given by 
	 \begin{equation}
	 	\begin{split}
	 		s^i(y^i, x) = - C^i_l  +  \lambda_i (  \sum_{j=1}^{m} h_{j,i} x^j y^i    
	 		- r_i B_u \log(a^i + (z^i)^\top y^i )),  \notag
	 	\end{split} 
	 \end{equation}
	 where $y^i$  is the transmission power of user $ i $ and satisfies $0 \leq \underline{l}^i \leq   y^i \leq \bar{l}^i  $, 
	 $ C^i_l \in \mathbb{R} $ is date transmission capacity, $ \lambda_i $ is the allocation parameter, 
	 $ h_{j,i} $ is the channel gain from operator $ j $ to user $ i $, 
	 $r_i$ is the revenue that user $i$ gains for unit data rate transmitted, and 
	 $B_u$ is the data size in the unlicensed spectrum.   
	 Moreover,  the cost function of each operator $ j \in \mathcal{I}_L $ is given by 
	 \begin{flalign} 
	 	\theta^j (x, y) = -x^j ( \sum_{i=1}^{N} h_{j,i} \lambda^i y^i ), \notag
	 \end{flalign}
	 where $ x^j > 0 $ is the price of operator $ j $ in the unlicensed spectrum. 
	 \par 
	As depicted in Fig.\ref{WIFI}, operators maintain communication with their neighboring operators and establish connections with their respective users, resulting in a clustered information structure in this case.  
	 \par Consequently, the two examples mentioned above can be effectively modeled using networked MLMF Stackelberg games featuring a clustered information structure.
\section{Algorithm Design}\label{section3} 
\par In this section, we propose Algorithm \ref{algorithm1} for SE seeking of a networked MLMF Stackelberg game with clustered information structure as depicted in Fig.\ref{clustered}.
\subsection{The Implicit Differential Approximation Method of An MLMF Stackelberg Game}
\par To calculate the SE by gradient methods, we need to obtain the hyper-gradient of leader $ j\in \mathcal{I}_L $ as follows
\small
\begin{equation} \label{definitionofhypergradient}
		\nabla_{x^j} \Phi^j(x):=\nabla_{x^j} \theta^j(x, y) + \frac{\partial y^*(x)}{\partial x^j}\nabla_y \theta^j (x, y)|_{y=y^*(x)},
\end{equation}
\normalsize
where $\Phi^j(x):=\theta^j (x,  y^*(x) )$ is utilized to avoid ambiguity caused by differentiation, since $\theta^j (x,  y^*(x) )$ is a function w.r.t. $x$ while $\theta^j(x,y)$ is a function w.r.t $(x,y)$.
It reveals that leader $j\in \mathcal{I}_L$ needs  $\frac{\partial y^*(x)}{\partial x}$ to make an optimal decision, whereas $\frac{\partial y^*(x)}{\partial x^j}$ is a global information of followers. 
\par The approximation to $\frac{\partial y^*(x)}{\partial x^j}$ is revealed by the following lemma.
\begin{lemma} \label{lemma2}
	Suppose Assumptions  \ref{assumptionoffunction} and \ref{assumption_lipschitz} hold. 	
	For any $ j,h\in \mathcal{I}_L $,  $ i\in \mathcal{I}_F $,
	any $  x_k\in \mathbb{R}^q$, $ y_{T,k} \in  \mathbb{R}^{p}  $, 
	\small
	\begin{flalign} \label{lemma1}
		&\nabla_{x^j}y^{*}( x_k) = \text{col}( \frac{\partial y^{1,*}( x_k)}{\partial x^j}, \dots,  
		\frac{\partial y^{i,*}( x_k)}{\partial x^j}, \dots, 
		\frac{\partial y^{N,*}( x_k)}{\partial x^j}             )  \notag \\
		& =-E^j \, \, \text{diag}(  ( J^{j,*}_{\mathcal{P}_h,k}  )_{h\in \mathcal{I}_L}  )
		\left(
		\text{diag}
		(  (   H^{*}_{\mathcal{P}_h,k}  )_{h\in \mathcal{I}_L}  )
		\right)^{-1}, 
	\end{flalign}
	\normalsize
	where $ E^j  =[I_{q_j}, \dots, I_{q_j}, \dots, I_{q_j}] \in  \mathbb{R}^{q_j\times Nq_j}$,
	$ J^{j,*}_{\mathcal{P}_h,k} := \text{diag}(\nabla_{x^j}\nabla_{z}s^i(z ,x_k)_{i\in \mathcal{P}_h})|_{z=y^{i,*}(x_k)}$, and
	$ H^{*}_{\mathcal{P}_h,k} := \text{diag}(\nabla_{z}^2s^i(z,x_k)_{i\in \mathcal{P}_h})|_{z=y^{i,*}(x_k)}$.
\end{lemma}
\par \textit{Proof:} See appendix \ref{proofoflemma2} for details.\hfill$\blacksquare $
\par Lemma \ref{lemma2} clarifies the relationship between global and local information in an MLMF Stackelberg game, offering a theoretical guarantee for the subsequent application of the network consensus approach.
\subsection{Notations for Global Information }
\par 
In this context, we use the superscript $*$ to indicate substituting the variable $y^{i}_{T,k}$ (or $y_{T,k}$)
with the optimal response function $y^{i,*}(x_k)$ (or $y^*(x_k)$) at the corresponding position at time $k$,
where $T$ is the iteration counts of follower $i$, see in (\ref{algorithmline1}).
Corresponding notations about 
$J^{j,*}_{\mathcal{P}_h,k} $ 
and $ H^{*}_{\mathcal{P}_h,k}$ in Lemma \ref{lemma2},
$J^j_{\mathcal{P}_h,k}$ and $H_{\mathcal{P}_h,k}$ with substituting $y^{i}_{T,k}$ are present as follows
\small
\begin{equation}
	\begin{aligned}
		&J^j_{\mathcal{P}_h,k} := \text{diag}(\nabla_{x^j}\nabla_{z}s^i(z,x_k)_{i\in \mathcal{P}_h})|_{z=y^i_{T,k}},	\\  \notag
		&H_{\mathcal{P}_h,k} :=  \text{diag}(\nabla^2_{z}s^i(z,x_k)_{i\in \mathcal{P}_h})|_{z=y^i_{T,k}}. 
	\end{aligned}
\end{equation}
\normalsize
\par It can be inferred from above that the  calculation of $\frac{\partial y^*(x) }{\partial x^j}$ requires global information about all followers.
However, with the communication limitation of clustered information structure,  it brings significant challenges of resolving.
\par The aforementioned variable definitions are all specific to the local information known by a particular leader (i.e., leader $j$). In order to analyze the equilibrium point of the whole system, it is necessary to define global variables, which are listed as follows:
\small
\begin{flalign}
& Z^{j,R}_{\mathcal{P}_h,k}:= J^j_{\mathcal{P}_h,k} (H_{\mathcal{P}_h,k})^{-1},   \notag \\
& Z^{R}_{\mathcal{P}_h,k}:= J_{\mathcal{P}_h,k} (\boldsymbol{I}_m \otimes H_{\mathcal{P}_h,k})^{-1},	\notag
\end{flalign}
\normalsize
where $J_{\mathcal{P}_h,k} := \text{diag}(({J^j_{\mathcal{P}_h,k}})_{j\in \mathcal{I}_L} )$.
Moreover, the stacked matrices are defined as 
$ J_{k}:= \text{diag}((J_{\mathcal{P}_h,k} )_{h\in \mathcal{I}_L}) $ and 
$ H_{k}:= \text{diag}((\boldsymbol{I}_m \otimes H_{\mathcal{P}_h,k} )_{h\in \mathcal{I}_L})  $, and we have 
\small
\begin{equation}
Z_k := J_k (H_k)^{-1}.
\end{equation}
\normalsize
We present Table \ref{table1} to clarify the notations about Jacobian, Hessian, and Jacobian-Hessian-Inverse (J-H-I) matrix with substituting variable $y_{T,k}$ or function $y^*(x_k)$. 
\small
\begin{table}[htbp]  
	\centering	
		\caption{ Substituting variable $y_{T,k}$ or function $y^*(x_k)$   into second order mappings of followers in cluster $j$  with fixed $T$ at time $k$  }\label{table1}
		\begin{tabular}{cccccc}
			\toprule 		
			&  \makecell[c]{ Jacobian }
			&\makecell[c]{ Hessian } 
			& \makecell[c]{ J-H-I  } 
			\cr 
			\midrule 
			$y_{T,k}$ & $ J_{\mathcal{P}_j,k} $ 
			& $ 	H_{\mathcal{P}_j,k}  $
			& $ 	Z^{R}_{\mathcal{P}_j,k}  $ 
			\cr 
			\midrule 
			$y^*(x_k)$ & $ J^*_{\mathcal{P}_j,k} $ 
			& $ 	H^*_{\mathcal{P}_j,k}  $
			& $ 	Z^{*}_{\mathcal{P}_j,k}  $ 
			\cr
			\bottomrule
	\end{tabular}
\end{table}
\normalsize
\subsection{Four Questions to Final Results} \label{4qs}
\par 
Consequently, it raises a final question: 
\par \textbf{ How to compute $\frac{\partial y^*(x) }{\partial x^j}$ distributively?}
\par To answer this final question and  overcome the obstacles posed by information deficiencies, we 
propose Algorithm \ref{algorithm1} guided by the following four fundamental questions.
\par \textbf{Q1: How to obtain $y^{i,*}(x_k)$, $J^*_{\mathcal{P}_j,k}$, and $H^*_{\mathcal{P}_j,k}$?} 
\par 
By Lemma \ref{lemma2}, we can deduce that both $J^{j,*}_{\mathcal{P}_h,k}$ and $H^{*}_{\mathcal{P}_h,k}$ rely on the value of $y^{i,*}(x_k)$. 
In the algorithm, the approximation of $y^{i,*}(x_k)$ is achieved by utilizing $T$ iterations of gradient descent (GD), obtained through (\ref{algorithmline1}), resulting in $y^i_{T,k}$.
To get a warm start, we let the output $y^i_{T,k-1 }$  equivalent to the initialization $ y^i_{0,k} $ of every GD. 
Subsequently, each follower $i$ computes $\nabla_{x^j}\nabla_{y^i}s^i(y^i_{T,k},x_k)$ and $\nabla_{y^i}^2s^i(y^i_{T,k},x_k)$ using a local second order oracle. These values are then transmitted to their corresponding leader $j$ to form approximations $J^{j}_{\mathcal{P}h,k}$ and $H_{\mathcal{P}h,k}$.
\par Invoking Lemma \ref{lemma2}, we can infer that the calculation of the local J-H-I $Z^{R}_{\mathcal{P}_j,k}$ is necessary.
To better illustrate the approximation methods in \textbf{Q2} and \textbf{Q3}, the optimizer and the estimator of leader $j \in \mathcal{I}_L$ are presented in Table \ref{table2}.
\small
	\begin{table}[H] 
			\centering	
			\caption{ The Optimizer and the estimator  for leader $j$  } \label{table2}
			\begin{tabular}{cccccc}
				\toprule 		
				& \makecell[c]{Leader $j$}  \\
				\midrule 
				Required value 
				& $Z^R_{\mathcal{P}_j,k}$ (not have) \\
				\midrule 
				Optimizer 
				& $Z_{\mathcal{P}_j,D,k} \rightarrow Z^R_{\mathcal{P}_j,k}$\\
				\midrule 
				Required global value 
				& $Z_{D,k} = \text{diag}(Z_{\mathcal{P}_j,D,k})_{j\in \mathcal{I}_L}$ (not have) \\
				\midrule 
				Estimator  
				& $\hat{Z}^{j}_{B,k} \rightarrow Z_{D,k}$ \\
				\bottomrule
		\end{tabular}
	\end{table}
	\normalsize 
\par \textbf{Q2:  How to approximate the local J-H-I value $Z^{R}_{\mathcal{P}j,k}$ ? } 
\par 
To obviate the computation burden posed by matrix inversion in  $ Z^{R}_{\mathcal{P}_j,k}:= J_{\mathcal{P}_j,k} (\boldsymbol{I}_m \otimes H_{\mathcal{P}_j,k})^{-1} $, we utilize 
$ Z_{\mathcal{P}_j,D,k} $ to approximate $ Z^{R}_{\mathcal{P}_j,k} $, which is obtained by solving a optimization problem as follows:
\small
\begin{equation} \label{minplm}
	\begin{aligned}
		\underset{ Z_{\mathcal{P}_j,k} } {\operatorname{min}}  \,
	    \frac{1}{2} \text{tr}(Z_{\mathcal{P}_j,k} ( \boldsymbol{I}_m \otimes H_{\mathcal{P}_j,k})  Z_{\mathcal{P}_j,k}^{\top})
	    - \text{tr}( J_{\mathcal{P}_j,k} Z_{\mathcal{P}_j,k}^\top ),
	\end{aligned}
\end{equation}
\normalsize
which utilizes $ D $ steps gradient decent starting from $ Z_{\mathcal{P}_j,0,k} = Z_{\mathcal{P}_j,D,k-1} $ to $ Z_{\mathcal{P}_j,D,k} $.
\par In conjunction with the definition of  $\nabla_{x^j} \Phi^j(x)$ and (\ref{algorithmline2}), we infer that leader $j$ requires not only local information $Z_{\mathcal{P}_j,D,k}$, but also global information $Z_{D,k}:= \text{diag}(Z_{\mathcal{P}_1,D,k},\dots, Z_{\mathcal{P}_h,D,k},\dots,Z_{\mathcal{P}_m,D,k})  $. 
\par \textbf{Q3:  How to approximate $ Z_{D,k} $ ? } 
\par To enable each leader to have the global knowledge $ Z_{D,k} $, an undirected graph among leaders is 
utilized.
We assume each leader $ j $ has an estimator $ \hat{Z}^{j}_{B,k}:=\text{diag} (\hat{Z}^{j(1)}_{B,k},\dots,\hat{Z}^{j(h)}_{B,k},\dots,\hat{Z}^{j(m)}_{B,k}) $ to estimate $Z_{D,k}$ by $B$ rounds communication to its leader neighbors,
where each item of $ \hat{Z}^{j}_{B,k} $ corresponds to the respective counterpart of $ Z_{D,k} $, i.e., $ \hat{Z}^{j(h)}_{B,k} $ approximates $ Z_{\mathcal{P}_h,D,k} $.
It is worth noting that the iteration method for solving (\ref{minplm}) has a smaller computational complexity compared to methods that are hybrid with matrix inversion if the dimension is large, see \cite{ghadimi2018approximation,blondel2000survey}.
\par By now, for leader $j$, the estimator  $ \hat{Z}^{j(h)}_{B,k} $ encompasses all the necessary information for computing the estimation of the hyper-gradient $\nabla_{x^j} \hat{\Phi}^j(x)$ and updating the decision variable. 
\par \textbf{Q4:  How to update the decision variable of each leader? } 
\par Note that not all information from the estimator $ \hat{Z}^{j(h)}_{B,k} $ is necessary, leader $j$ needs to extract the relevant information from the estimator.
In conjunction with the analysis of Step 1, 2, and (\ref{algorithmline2}), we infer that leader $j$ requires $ Z^{j,R}_k:=\text{diag}((Z^{j,R}_{\mathcal{P}_h,k})_{h\in\mathcal{I}_L})$, where each block item on the diagonal needs to be extracted from the information coming from the corresponding leader, for example, $Z^{j,R}_{\mathcal{P}_h,k}$ is approximated by    
$ Z^j_{\mathcal{P}_h,D,k} $ that is $ j $-th item on the  diagonal of $ Z_{\mathcal{P}_h,D,k}$,
furthermore, the $j$-th item of $ \hat{Z}^{j(h)}_{B,k} $ by (\ref{algorithmestimate}). 
\par We utilize $ R^j_h:= [0,\dots, I_{n_hq_j \times n_hq_j}, \dots, 0]  \in \mathbb{R}^{n_h q_j \times n_hq} $ and 
$ U^j_h= [0, \dots , I_{p_{\mathcal{P}_h} \times p_{\mathcal{P}_h}}, \dots, 0 ]^{\top} \in \mathbb{R}^{ m p_{\mathcal{P}_h} \times p_{\mathcal{P}_h}  }$ to extract $j$-th block item from $\hat{Z}^{j(h)}_{B,k}$ to approximate $Z^{j}_{\mathcal{P}_h,D,k} $, i.e.,  $ R^j_h \hat{Z}^{j(h)}_{B,k} U^j_h \rightarrow Z^{j}_{\mathcal{P}_h,D,k}  $,
which indicates the main design idea of (\ref{algorithmestimate2}) .
Subsequently, in conjunction with Lemma \ref{definitionofhypergradient} and  projected gradient method lead readily to (\ref{estimatehypergradient}) and (\ref{project}).
\begin{algorithm}  [htbp]
	\caption{Distributed SE Seeking for MLMF Stackelberg Games} \label{algorithm1}
	\textbf{Initialize}:
For all  $  j,h \in \mathcal{I}_L  $, $ i,l \in \mathcal{I}_F $:
$y^i_{0,0} \in \mathbb{R}^{p_i} $,
$ v^j_{\mathcal{P}_h,0,0} $,
$ J^j_{\mathcal{P}_h,0} $,
$ \hat{Z}^{j(h)}_{0,0} $,
input $ T $, $ D $, and $ B $, 
 \begin{algorithmic}[1]
 	\State  {set $ x^j_{0}\in \Omega_j \subseteq \mathbb{R}^{q_j}$, $k=1$.}
 	\Repeat
 	\For{each leader $ j \in \mathcal{I}_L $}
 		\begin{flushleft}
 			\textbf{\text{\bfseries Q1}. Approximation of $y^{i,*}(x)$, $J_{\mathcal{P}_j,k}$, and $H_{\mathcal{P}_j,k}$}\notag	
 		\end{flushleft}
 	\State \hspace{-3em} for any $ i \in \mathcal{I}_F $ and 
 		set $ y^{i}_{0,k}=y^{i}_{T,k-1} $, if $t\leq T$ updates:
 	\begin{equation} \label{algorithmline1}
 		\hspace{0.5cm}	y^i_{t,k} = y^i_{t-1,k} - \alpha \nabla_{y^i} s^i (y^i_{t-1,k}, x_k),
 	\end{equation}
	$y_{T,k} $ is obtained by $y_{T,k} \leftarrow 	 \text{col}(y^{1}_{T,k},\dots, y^{j}_{T,k},\dots, y^{m}_{T,k})),$
	\State \hspace{-3em} collects  information from corresponding cluster:
	\begin{flalign}
		J_{\mathcal{P}_j,k} &\leftarrow 
		\text{diag}(\nabla_{x^j}\nabla_{y^i} s^i(y^i_{T,k},x_k)_{i\in \mathcal{P}_j,j\in \mathcal{I}_L}), \\
		H_{\mathcal{P}_j,k} &\leftarrow \text{diag}(\nabla_{y^i}^2s^i(y^i_{T,k},x_k)_{i\in \mathcal{P}_j}),
	\end{flalign}
	 \begin{flushleft}
		\textbf{\text{\bfseries Q2}. Approximation of Local J-H-I for leader $j$} \notag
	\end{flushleft}
    \State \hspace{-3em}
    set  $ Z_{\mathcal{P}_j,0,k}=Z_{\mathcal{P}_j,D,k-1} $, if $ d\leq D $ updates:
    \begin{flalign} \label{algorithmline2}
    	Z_{\mathcal{P}_j,d,k}  &= Z_{\mathcal{P}_j,d-1,k}
    	-\gamma  Y_{\mathcal{P}_j,d-1,k},   \notag \\
    	Y_{\mathcal{P}_j,d-1,k} &= Z_{\mathcal{P}_j,d-1,k} (\boldsymbol{I}_m \otimes H_{\mathcal{P}_j,k}) - J_{\mathcal{P}_j,k},
    \end{flalign}
    \begin{flushleft}
    	\textbf{\text{\bfseries Q3}. Approximation of  $ Z_{D,k} $ for leader $j$} \notag
    \end{flushleft}
    \State \hspace{-3em} for every $h\in \mathcal{I}_L $,  if $b\leq B$ updates:
    \small
    \begin{flalign}
    	&\hat{Z}^{j(h)}_{b,k} = (\sum\nolimits_{g\in \mathcal{N}_j}w^{jg} \hat{Z}^{g(h)}_{b-1,k}) 
    	+ \xi^j_h w^{jh} (Z_{ \mathcal{P}_h,D,k}
    	-\hat{Z}^{j(h)}_{b-1,k}),   \label{algorithmestimate}\\ 
    	&\hat{Z}^{j,RU}_{B,k}  \leftarrow \text{diag}(
    	R^j_1 \hat{Z}^{j(1)}_{B,k}U^j_1 ,\dots,R^j_h \hat{Z}^{j(h)}_{B,k} U^j_h ,\dots,R^j_m  \hat{Z}^{j(m)}_{B,k}U^j_m), \label{algorithmestimate2}
    \end{flalign} 
    \normalsize
    \begin{flushleft}
    	  {\textbf{\text{\bfseries Q4}. Update the decision variable  for leader $j$}} \notag
    \end{flushleft}
    \normalsize
    \State \hspace{-3em}{ update $x^j_{k+1} $ after calculating  $\nabla_{x^j}  \hat{\Phi}^j (x_{k})$:}
   	\small\begin{flalign} \label{estimatehypergradient}
    	&\nabla_{x^j}  \hat{\Phi}^j (x_k) = \nabla_{x^j} \theta^j (x^j_k,x^{-j}_k,y_{T,k}) \notag \\ 
    	& \qquad \qquad \qquad \qquad
    	-E^j \hat{Z}^{j,RU}_{B,k} \nabla_{y} \theta^j (x^j_k,x^{-j}_k,y_{T,k}),       	
    \end{flalign}
    \normalsize
        \small
        \begin{flalign}\label{project}
        &x^j_{k+1}  = \Pi_{\Omega_j} (x^j_k  -  \beta_k \nabla_{x^j}  \hat{\Phi}^j (x_k)),	
        \end{flalign}
        \normalsize
 	\EndFor
 	\State { $k  =  k  + 1 ,$}
 	\Until{end}	
 \end{algorithmic}
\end{algorithm}
\par The algorithm is shown as Algorithm \ref{algorithm1}, where $ \xi^j_h $ in (\ref{algorithmestimate}) is a positive constant fulfilling $ 0<\xi^j_h <\frac{w^{jj}}{w^{jh}}$ in matrix $ \tilde{W}^ h,\, \forall w^{jh} \neq 0,\,\forall  h\in \mathcal{I}_L$.
The matrix $\tilde{W}^h $  is  $W^h$ with diagonal entries replaced by $[\tilde{W}^h]_{rr}= [W^h]_{rr}- \xi^{r}_h [W^h]_{rh}$, $r\in \mathcal{I}_L$.
\begin{figure}[htbp] 
	\centering
	\includegraphics[width=3.6in]{ 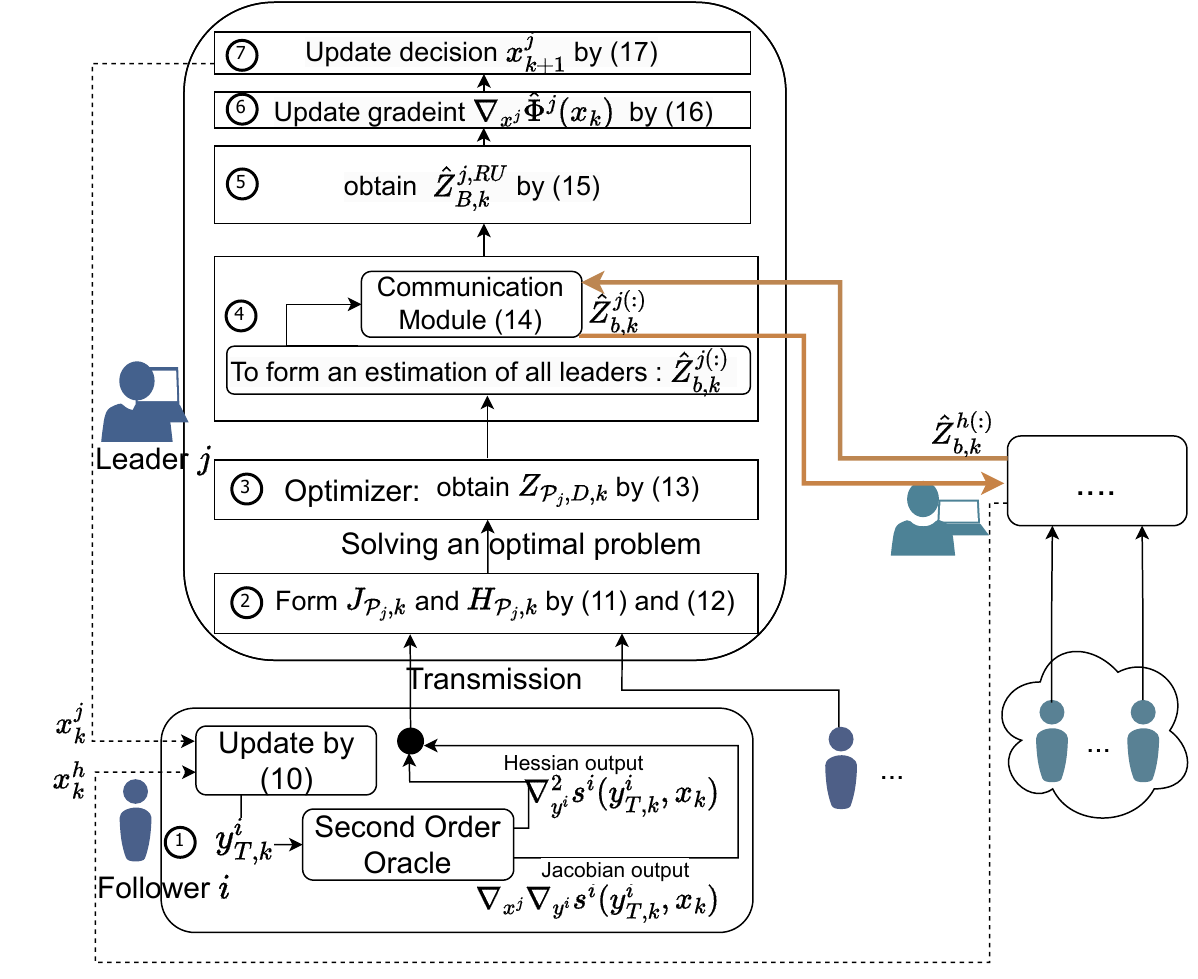}\\
	\caption{The flowchart of Algorithm 1. }
\end{figure}
\section{Main Results} \label{section4}
\par In this section, we present the main results of an MLMF Stackelberg games with a clustered information structure through approximate analysis and convergence analysis.  
\par The theoretical result of the convergence analysis relies on three approximation lemmas from the approximate analysis in Subsection \ref{approx}, corresponding to \textbf{Q1}, \textbf{Q2}, and \textbf{Q3} in Subsection \ref{4qs}, respectively. Furthermore, an equilibrium analysis of the competition between leaders is introduced following the subsequent discussions on pseudo-gradient and a corresponding VI condition.  
Then, the pseudo-gradient of all leaders is defined as
\small 
\begin{flalign}\label{pseudogradient}
	\Psi(x ,y^*(x )):= \text{col}(  ( \nabla_{x^j}\Phi^j(x))_{j\in \mathcal{I}_L}  ).
\end{flalign}
\normalsize
 Moreover, we introduce a lemma of the optimal condition of VI as follows:
\begin{lemma} (\cite{facchinei2007finite}, Proposition 1.4.2)  \label{lemmaofoptimalcondition}
	Suppose Assumptions \ref{assumptionoffunction} and \ref{assumption_lipschitz} hold,
 the solution of a variational inequality problem $ \text{VI}(\Omega, \Psi) $, i.e., 
	$ (x-x^\diamond)^T\Psi(x^\diamond)\geq0 $, $ \forall x\in \Omega $ is equivalent to 
	$	x^\diamond = \prod_{\Omega}(x^\diamond -  \beta  \Psi(x^\diamond)  )$.
\end{lemma}
\par Next, we impose a mild assumption on the uniqueness of the Nash equilibrium of leaders. 
\begin{assumption} \label{assumptionofstrictmonotone}
	The pseudo-gradient $ \Psi(x)  $ is strictly monotone on $ \Omega $, for any $ x, \, \bar{x} \in \Omega $, 
	$ \left\langle  \Psi(x) - \Psi(\bar{x})       ,   x- \bar{x} \right\rangle   >0$. 
\end{assumption}
The above strictly monotone assumption guarantees the uniqueness of the equilibrium,  which is common in many literature, e.g., \cite{belgioioso2020distributed}.  	
\subsection{Approximate Analysis} \label{approx}
\par In this subsection, we introduce some lemmas about approximate method. 
\par Foremost, we analyze the estimate method for hyper-gradient of each leader  $ j\in \mathcal{I}_L $ in the following. 
\begin{lemma}[The hyper-pseudo-gradient] \label{lemmaofhypergradient}
	   The hyper-gradient of leader $ j \in \mathcal{I}_L$ at iteration $ k $ is given by $	\nabla_{x^j} \Phi^j(x_k)
	:=\nabla_{x^j} \theta^j(x_k, y) 
	-E^j R^j Z^*_k U^j  \nabla_{y} \theta^j (x_k,y)|_{y=y^*(x_k)} $.
 Furthermore, the hyper-pseudo-gradient of leaders $ \Psi^*(x_k):=\Psi(x_k ,y^*(x_k)) $ is given by 
\small
\begin{flalign} \label{lemmaline1}
	\Psi^*(x_k)=\Theta^*_x (x_k) - E R (\boldsymbol{I}_m \otimes Z^*_k)U \Theta^*_y (x_k),   
\end{flalign}
\normalsize
where $ \Theta^*_x (x_k) := \text{col}( \nabla_{x^j} \theta^j(x_k, y)_{j\in \mathcal{I}_L})|_{y=y^*(x_k)}  $, 
$ E:=\text{diag}((E^j)_{j\in \mathcal{I}_L}) $, $ R:=\text{diag}((R^j)_{j\in \mathcal{I}_L}) $, and
$ \Theta^*_y (x_k) := \text{col}( \nabla_{y} \theta^j(x_k, y)_{j\in \mathcal{I}_L})|_{y=y^*(x_k)} $.
\end{lemma}
\par \textit{Proof:} See appendix \ref{proofoftheorem2} for details.\hfill$\blacksquare $
\par Next, we introduce three lemmas corresponding to the three approximation steps introduced in section \ref{section3}. More proof details of Lemma \ref{lemmaofyTtoyx}, \ref{lemmaofinverse}, and \ref{lemmaofZ} can be found in the Supplementary Materials due to space limitations.	
\begin{lemma}[Best responses approximation](Theorem 2.1.12, \cite{nesterov2003introductory}) \label{lemmaofyTtoyx}
	Suppose Assumptions  \ref{assumptionoffunction} and \ref{assumption_lipschitz} hold. 
	If the step size fulfills $ 0< \alpha \leq \frac{2}{\mu + \ell_{ s,1}}$, the following inequality holds:
	\begin{flalign}	
		 \|  y_{T,k}   -   y^*(x_k)  \|^2 
		 \leq (1-   2 \alpha\Gamma)^T  \|  y_{0,k} - y^*(x_k)\|^2,
	\end{flalign}
 where $ \Gamma := \frac{ \mu \ell_{s,1}}{\mu + \ell_{ s,1}} $.  
\end{lemma}
\begin{lemma}[Local J-H-I approximation] \label{lemmaofinverse}
	Suppose Assumptions \ref{assumptionoffunction} and \ref{assumption_lipschitz} hold. 
	If $ \frac{1}{\mu}-\frac{1}{\sqrt{m p_M } \mu}  <  \gamma \leq \frac{1}{\ell_{ s,2 }} $, then the following inequality holds:
	\small
	\begin{flalign} \label{lemma-8-eq1}
		&\|    Z_{\mathcal{P}_j,D,k}   - Z^{R}_{\mathcal{P}_j,k} \|^2_F \notag \\
		& \qquad \leq  ( \sqrt{m p_{ \tiny P_j}} - \sqrt{m p_{ \tiny P_j}} \gamma \mu )^{2D} 
		 \|    Z_{\mathcal{P}_j,0,k}   - Z^{R}_{\mathcal{P}_j,k } \|^2_F , 
	\end{flalign}	 
\normalsize
where $  p_{ \tiny P_j} $ is the dimension of the collective decision of $ j $-cluster followers, and $ p_M  $ is the maximum of set  $\{  p_{ \tiny P_1} , \dots,  p_{ \tiny P_j} , \dots , p_{ \tiny P_m} \}  $ .  
Furthermore, 
\small
\begin{flalign}
	\|  Z_{D,k}  - Z^R_{k}  \|^2_F \leq 
	(m\sqrt{p_M}-m\sqrt{p_M}\gamma\mu )^{2D} 
	\| Z_{0,k}  - Z^R_k      \|^2_F
\end{flalign}
\normalsize
where $ Z_{D,k} := \text{diag}( (  Z_{\mathcal{P}_j,D,k } )_{j\in \mathcal{I}_L}   ) $.
\end{lemma}
\par \textit{Proof:} See appendix \ref{proofoflemmaofinverse} for details.\hfill$\blacksquare $
\par Recall that each leader $ j\in \mathcal{I}_L $ can only have the second order information of followers within the corresponding cluster due to the aforementioned communication limitations. 
To obtain the global information, 
leader $ j $ utilizes an estimator $ \hat{Z}^j_{B,k} $ to approximate the collective $ Z_{D,k}  $.
\par To better explain the following lemma, we first introduce some notation.
We denote by $ \tilde{\boldsymbol{W}}^h : = \tilde{W}^h \otimes I_{n_h q } $
$ \tilde{\boldsymbol{W}} :=  \text{diag} (  ( \tilde{\boldsymbol{W}}^h      )_{ h\in \mathcal{I}_L}     )    $,
$ \hat{Z}^{:{(h)}}_{b,k} := \text{col}  (({  \hat{Z}^{j({h})}_{b,k}  } )_{j\in \mathcal{I}_L})  $, 
$ \hat{Z}^{:(:)}_{b,k} := \text{diag} ((\hat{Z}^{:(h)}_{b,k})_{h\in \mathcal{I}_L}) $,
and  
$ Z_{:,D,k}  := \text{diag}  ( (   \boldsymbol{1}_m \otimes  Z_{\mathcal{P}_h , D, k   })_{h\in \mathcal{I}_L}   )$. 
\begin{lemma} [Global followers information approximation] \label{lemmaofZ} 
	 Leaders' estimators to leader $ h\in \mathcal{I}_L $ have to satisfy the following equality
\small
	 \begin{flalign}
	 	\hat{Z}^{:{(h)}}_{b,k} - \boldsymbol{1}_m \otimes Z_{\mathcal{P}_h,D,k}   
	 	= \tilde{\boldsymbol{W}}^{h} (\hat{Z}^{:{(h)}}_{b-1,k}   -  \boldsymbol{1}_m \otimes  Z_{\mathcal{P}_h,D,k}).
	 \end{flalign}	
\normalsize
 Additionally, the collective estimators satisfy 
 \small
 \begin{flalign}
 	\hat{Z}^{:(:)}_{b,k} -  Z_{:,D,k} =  \tilde{\boldsymbol{W}} 
 	( \hat{Z}^{:(:)}_{b-1,k}  -   Z_{:,D,k}  ).
 \end{flalign}
\normalsize
There exists a norm $ \|\cdot\|_w $ satisfying the following equality
\small
\begin{flalign}
	\| \hat{Z}^{:(:)}_{b,k} -  Z_{:,D,k}     \|_w \leq \sigma_w \|  \hat{Z}^{:(:)}_{b-1,k}  -   Z_{:,D,k}   \|_w,
\end{flalign}
\normalsize
where $ \sigma_w <1 $. 
\end{lemma}
\par \textit{Proof:} See appendix \ref{proofoflemma10} for details.\hfill$\blacksquare $
\par Since $  \tilde{\boldsymbol{W}} $ is symmetric, we can choose $ 2 $-norm directly, i.e., 
$ 	\| \hat{Z}^{:(:)}_{b,k} -  Z_{:,D,k}     \| \leq  \sigma_2 \|  \hat{Z}^{:(:)}_{b-1,k}  -   Z_{:,D,k}   \| $, 
where $  \sigma_2:=\rho (\tilde{\boldsymbol{W}}) $.
Denote by  $ \hat{Z}_{B,k}:= \text{diag}(  (  \hat{Z}^j_{B,k}  )_{j\in \mathcal{I}_L}  ) $,  
and in conjunction with  the definition of $ \hat{Z}^{:(:)}_{b,k} $, $ Z_{:,D,k} $, and $ Z_{D,k}$, it can be readily verified that 
$ \| \hat{Z}^{:(:)}_{b,k} - Z_{:,D,k}    \|_F  =  \|   \hat{Z}_{B,k}    - \boldsymbol{I}_m \otimes   Z_{D,k}   \|_F  $. 
In light of Lemma 6 in \cite{pu2020push},  for $ W \in \mathbb{R}^{n \times n}$ of rank $ r $, the following inequalities hold: 
\begin{flalign}
	 \| W \|_2  \leq \|W\|_F,   \quad  \|W\|_F  \leq   \sqrt{r} \|W\|_2.  \notag 
\end{flalign}   
Thus, we have the following inequality
\begin{equation}
	\|   \hat{Z}_{B,k}    - \boldsymbol{I}_m \otimes   Z_{D,k}   \|_F  \leq   \sqrt{r_z}  \sigma_2^B \|\hat{Z}_{0,k}   - \boldsymbol{I}_m \otimes   Z_{D,k}   \|_F,
\end{equation}
where $ r_z $ is the rank of $ (\hat{Z}^{:(:)}_{b,k} - Z_{:,D,k} ) $.
\subsection{Convergence Analysis}
\par In this subsection, we discuss the convergence results under outer loop diminishing and constant step-size , respectively. Next, we present some lemmas regarding the distance between the iteration point and the equilibrium point.
\subsubsection{Diminishing Step-size}
\par For brevity of notation, we utilize the following symbols to replace some complex expressions:
\small
\begin{flalign}\label{notation}
		&C_T   :=   (1-2 \alpha\Gamma)^T,\,\,   C_D:=  (m\sqrt{p_M}-m\sqrt{p_M}\gamma\mu )^{2D}, \notag\\
		&C_B   :=  r_z \sigma_2^{2B},\,\notag\\
		& \Xi_1 \sim  \Xi_8 \,\,\text{are}  \,\,\text{constants}
		\,\,\text{defined}  \,\,\text{in}  
		\,\,\text{appendix}\,\,\ref{constantnotations}, \\
		&C:=\Xi_5 C_T + 18mN^2\ell^2_{\theta,0}C_D + 18N^2\ell^2_{\theta,0}C_B, \notag 
\end{flalign}
\normalsize
where  $ C_T $, $ C_D $, and $ C_B $ are variables whose values depend on iteration counts, step sizes and spectrum radius, i.e., $ T $, $ D $, $ B $, $ \alpha $, $ \gamma $, and $ \sigma_2 $.  Furthermore, we introduce additional notations 
from $ \Xi_1  $ to $ \Xi_8  $
to represent constants that arise in the analysis. 
The specific forms of these constants are provided in the appendix \ref{constantnotations} .

\begin{lemma} [Distance to equilibrium point] \label{lemmaofdistanceofz}
\par 	Suppose Assumptions \ref{assumptionoffunction} and \ref{assumption_lipschitz} hold. Then, we have
\small
\begin{equation}
	\begin{split}
		&\| \hat{Z}_{B,k}  -  \boldsymbol{I}_m \otimes Z^*(x^{\diamond})    \|^2_F  \\
		& \leq  	\Xi_2 \| x_k -x^\diamond  \|^2 +\Xi_1 C_T  \| y_{0,k}- y^*(x_k) \|^2 \\ 
		& + 6mC_D \| Z_{0,k}  - Z^R_k      \|^2_F  + 6C_B \|\hat{Z}_{0,k}   - \boldsymbol{I}_m \otimes   Z_{D,k}   \|^2_F,   \notag \\ 
	\end{split}
\end{equation}
\normalsize
moreover, 
\small
\begin{equation}
	\begin{split}
		&\| \hat{\Psi}(x_{k}) -  \Psi(x^\diamond)    \|^2  \leq \Xi_4 \|x_{k}-x^\diamond \|^2 +  \Xi_5 C_T \| y_{0,k}- y^*(x_k) \|^2 \\
		& + 18 N^2\ell^2_{\theta,0} (mC_D\| Z_{0,k}  - Z^R_k      \|^2_F + C_B\|\hat{Z}_{0,k}   - \boldsymbol{I}_m \otimes   Z_{D,k}   \|^2_F), \notag \\
	\end{split}
\end{equation}
\normalsize
where $ x^\diamond $ is the equilibrium point, $ \Xi_1 $, $ \Xi_2 $,  $ \Xi_4 $, $ \Xi_5 $, $ C_T $, $ C_D $, and $ C_B $ are defined in (\ref{notation}).  
\end{lemma}
\par \textit{Proof:} See appendix \ref{proofoflemmaofdistanceofz} for details.\hfill$\blacksquare $
\begin{proposition} \label{lemmaofx0z0zhat0}
 Suppose Assumptions \ref{assumptionoffunction} and \ref{assumption_lipschitz} hold.
 If the $ T $, $ D $, and $ B $ fulfills the following inequality:
 \small
 \begin{equation} \label{iterationcounts}
 	\begin{split}
 		 &  T \geq \frac{\ln({2\pi +\frac{16\pi \Xi_3}{m} + \frac{128\pi \Xi_1}{3} +2 \beta^2_M \pi  \Xi_5 \Xi_8 })  }{\ln (\frac{1}{1-2 \alpha \Gamma})},   \\
 		 &  D \geq \frac{ \ln({ 2\pi+16m\pi+36 m N^2 \pi \beta^2_M \ell^2_{\theta,0} \Xi_8  })         }{  2\ln ( \frac{1}{  (m\sqrt{p_M}-m\sqrt{p_M}\gamma\mu )   }  )     },   \\
 		 &  B \geq  \frac{ \ln ( {   2\pi  + 36 N^2 \pi\beta^2_M \ell^2_{\theta,0} \Xi_8        }    )    }{2 \ln ({\frac{1}{r_z\sigma_2}})},
 	\end{split}
 \end{equation}
 \normalsize
where $  \pi> 1 $, $ \beta_M $ is the maximum of $ \{ \beta_k  \}^\infty_0 $ that fulfills $ 0 \leq \beta_{k} \leq \beta_{k-1} \leq 1, \forall k \geq 0    $,  $ \sum_{k=0}^{\infty} \beta_k  = \infty $ and $\sum_{k=0}^{\infty} \beta^2_k  < \infty  $,  $ \Xi_3 $, $ \Xi_5 $, and $ \Xi_8 $ are defined in (\ref{notation}). Then, we have 
\small
\begin{equation}
	\begin{split}
    \Delta_{k}  & \leq  (\frac{1}{\pi})^k \Delta_{0} + 2\beta_k^2 \sum_{s=0}^{k-1} (\frac{1}{\pi})^{s} \Xi_6 \Xi_8,   \\ 
		& \leq (\frac{1}{\pi})^k \Delta_{0} + \frac{2\beta_k^2 \Xi_6\Xi_8 \pi}{\pi-1} +  \frac{2\beta_k^2 \Xi_6\Xi_8 \pi}{\pi-1} (\frac{1}{\pi})^{k-1}, 
	\end{split}
\end{equation}
 \normalsize 
 where $ \Delta_{k}:= \| y_{0,k}- y^*(x_k) \|^2  +   \| Z_{0,k}  - Z^R_k \|^2_F + \| \hat{Z}_{0,k}  - \boldsymbol{I}_m \otimes Z_{D,k}   \|^2_F $,
  and $ \Delta_{0}:= \Delta_{k} $ when $ k=0 $.
\end{proposition}
 \par \textit{Proof:} See appendix \ref{proofoflemmaofx0z0zhat0} for details.\hfill$\blacksquare $
\par 
	By substituting the constants defined in (\ref{notation}) into the inequalities in (\ref{iterationcounts}), it can be inferred that $T$, $D$, and $B$ increase with the growing number of leaders $m$ or followers $N$, which in turn imposes a decrease in the convergence rate of Algorithm \ref{algorithm1}.    
\begin{remark}
	The step-size sequence  $ \{  \beta_k \}^\infty_0 $ is common in many research literature, such as \cite{lei2021distributed}. For instance, the sequence can be chosen by $ \beta_k = (k+1)^{-b} $, where $ b \in (1/2, 1] $.
	Furthermore, $ \beta_M $ that defined in Lemma \ref{lemmaofx0z0zhat0} can be readily established, 
	since the sequence is a decrease sequence in this case. 
\end{remark}
\begin{theorem} \label{theorem1}
        Suppose Assumptions \ref{assumptionoffunction}, \ref{assumption_lipschitz}, and \ref{assumptionofstrictmonotone} hold.
        If 
        $ 0< \alpha \leq \frac{2}{\mu + \ell_{ s,1}}$,  
        $ \frac{1}{\mu}-\frac{1}{\sqrt{m p_M } \mu}  <  \gamma \leq \frac{1}{\ell_{ s,2 }} $,
        $ 0 < \xi^j_h < \frac{w^{jj}}{w^{jh}} $, 
        the step-size sequence $ \{\beta_k\} _0 ^ \infty $ fulfills $ 0 \leq \beta_{k} \leq \beta_{k-1} \leq 1, \forall k \geq 0    $, 
        $ \sum_{k=0}^{\infty} \beta_k  = \infty $, $\sum_{k=0}^{\infty} \beta^2_k  < \infty  $,
        and the iteration counts $ T $, $ D $, and $ B $ satisfy the  (\ref{iterationcounts}) in Lemma \ref{lemmaofx0z0zhat0}, 
        then the sequences $ \{ x_k \}_0^\infty  $ and $ \{ y_{T,k} \}_0^\infty   $ 
        generated by Algorithm \ref{algorithm1} converge to $ x^\diamond $ and $ y^\diamond $, 
        which are the solution of (\ref{followergame}) and (\ref{leaderproblem}) respectively.
         Furthermore,
        $ (x^\diamond, y^\diamond) $ is a SE point of the $ m$-leader and $ N $-follower Stackelberg game (\ref{gamemodel}).       
\end{theorem}
\par \textit{Proof:} See appendix \ref{proofoftheorem1} for details.\hfill$\blacksquare $
\subsubsection{Constant Step-size}
\par Next, we discuss a constant step-size scenario, i.e., $ \beta_k =\beta $, 
where we utilize the strongly monotone assumption instead of strictly monotone. 
Then we have the following assumption:  
\begin{assumption}	\label{assumptionofstronglymonotone}
    The pseudo-gradient $ \Psi(x)  $ is strongly monotone with $ m_{\theta}>0 $ on $ \mathbb{R}^{q} $, for any $ x, \, \bar{x} \in \mathbb{R}^q $, 
	$ \left\langle  \Psi(x) - \Psi(\bar{x}),   x- \bar{x} \right\rangle   \geq m_{\theta} \| x-\bar{x}  \|^2 $. 
\end{assumption}
Additionally, let $ \Omega = \mathbb{R}^q $ hold, the optimal solution in Lemma \ref{lemmaofoptimalcondition} is equivalent to  $\Psi(x^\diamond) =0$.
\par Based on Lemma 2 in \cite{ji2021bilevel}, we have $	\|  \nabla_{x^j} \Phi^j (x^j, x^{-j}) -  \nabla_{x^j}  \Phi^j(\bar{x}^{j}, x^{-j})       \| \leq \ell_{ \Phi}  \|    x^j   -   \bar{x}^{j}                 \|,$  
for any $ j\in \mathcal{I}_L $, $ x^j$, $ \bar{x}^{j} \in \mathbb{R}^{q_j}$, where $\ell_{\Phi}$ 
is a constant. 
Moreover,  $\| \Psi(x)    -  \Psi(\bar{x})     \| \leq \ell_{\Phi} \| x -\bar{x}  \|$ is hold, for any $ x,\bar{x} \in \mathbb{R}^q$.
\begin{lemma} \label{lemma_of_lipsmono}
		Suppose  Assumptions \ref{assumptionofstronglymonotone} hold. In conjunction with the Lipschitz continuous property of the pseudo-gradient,
     	the following inequality holds :
	     $ \|x-\beta \Psi (x) - (\bar{x}-\beta \Psi(\bar{x}))  \|
		\leq \sqrt{1-2 m_{\theta} \beta + \ell_{\Phi}^2 \beta^2} \|  x  -  \bar{x}    \|	$, for all $ x,\bar{x} \in \mathbb{R}^n $,
		where $ \beta \in (0, \frac{2 m_{\theta}}{\ell_{\Phi}^2}) $.
\end{lemma}
\par For the notational convenience, we denote 
\small
\begin{equation} \label{notation2}
	\begin{split}
		&   \tilde{C}_T : =\sqrt{\Xi_5 C_T} , \,\, \tilde{C}_D:=  \sqrt{18  mN^2\ell^2_{\theta,0}C_D}   \\
		&   \tilde{C}_B:= \sqrt{18N^2\ell^2_{\theta,0}C_B}, \,\, G(\beta):= \sqrt{1-2m_{\theta}\beta + \ell_{\Phi}\beta^2}. \\
		&   \tilde{\boldsymbol{\varDelta}}_{k} := (\|  x_k  -  x^\diamond\| ,\| y_{0,k}- y^*(x_k) \| ,    \| Z_{0,k}  - Z^R_k \|_F,  \\
		&  \|  \hat{Z}_{0,k}  -\boldsymbol{I}_m \otimes Z_{D,k}   \|_F ).   
	\end{split}
\end{equation}
\normalsize
\begin{proposition} \label{proposition1}
		Suppose Assumptions \ref{assumptionoffunction}, \ref{assumption_lipschitz}, and \ref{assumptionofstronglymonotone} hold.
		If  $ x^\diamond $ fulfills the optimal solution in Lemma \ref{lemmaofoptimalcondition}, then we have the following linear system of inequality:
		\small
		\begin{equation}
        \tilde{\boldsymbol{\varDelta}}_{k+1}     \leq  \boldsymbol{M}(\beta)   \tilde{\boldsymbol{\varDelta}}_{k},
		\end{equation}
	\normalsize
	where  $  \boldsymbol{M}(\beta)= [m_{ij}]_{4\times4} , \forall i, j \in \{1,2,3,4\} $  is  a matrix and each row is given by 
	\small
	\begin{equation}
		\begin{split}
			  [m_{1 j}]_{1\times4} = & [ G(\beta) , \,\, \beta \tilde{C}_T ,\, \, \beta\tilde{C}_D   ,\, \,    \beta\tilde{C}_B                    ] , \\ \notag
			  [m_{2 j}]_{1\times4} = & [ \kappa\sqrt{2N\Xi_4}\beta , \, \sqrt{2C_T} + \kappa\sqrt{2N}\tilde{C}_T\beta 
			  ,\, \kappa\sqrt{2N}\beta\tilde{C}_D,     \\
			  & \, \kappa\sqrt{2N} \beta \tilde{C}_B   ],\\
			  [m_{3 j}]_{1\times4} = & [ \sqrt{\Xi_4 \Xi_7} \beta ,  \sqrt{\Xi_7} \tilde{C}_T \beta + 4 \sqrt{\Xi_3 C_T / m}, \\   
			  &\sqrt{\Xi_7}\tilde{C}_D\beta + \sqrt{2C_D},  \, \sqrt{\Xi_7} \beta \tilde{C}_B], \\
			  [m_{4 j}]_{1\times4} = & [ 2\sqrt{2m\Xi_4 \Xi_7}\beta ,  8\sqrt{2\Xi_3C_T} + 2\sqrt{2m\Xi_7}\beta\tilde{C}_T
			  ,     \\ 
			  & 4\sqrt{mC_D} +2 \sqrt{2m\Xi_7} \beta \tilde{C}_D , \, \sqrt{2C_B} + 2\sqrt{2m\Xi_7} \beta \tilde{C}_B ].
		\end{split}
	\end{equation}
   \normalsize
\end{proposition}
\par \textit{Proof:} See appendix \ref{proof_ofproposition1} for details.\hfill$\blacksquare $

\begin{theorem} \label{theorem2}	 
	 Suppose Assumptions \ref{assumptionoffunction}, \ref{assumption_lipschitz}, and \ref{assumptionofstrictmonotone} hold.
	 There exists a positive step-size $ \beta $ near $ 0 $,  which results in $ \boldsymbol{M}(\beta) <1 $ holds. 
	 The bound of $ \beta $ is given by $   0< \beta < \text{min}\{ \beta_{s}, \frac{2 m_{\theta}}{\ell_{\Phi}^2} \} $, where 
	 $ \beta_s $ is the smallest positive root of $ \text{det} (  I_4   -\boldsymbol{M}(\beta) ) =0 $.
	 With $\alpha$, $\gamma $, $\beta $, $ T $, $ D $, and $ B $ satisfy the same conditions 
	 as  in Theorem \ref{theorem1},  the sequences $ \{ x_k \}_0^\infty  $ and $\{y_{T,k}\}_0^\infty$ generated by Algorithm \ref{algorithm1} converge to $ x^\diamond $ and $ y^\diamond $, which $ (x^\diamond, y^\diamond) $ is a SE point of the $ m$-leader and $ N $-follower Stackelberg game (\ref{gamemodel}).
\end{theorem}
\par \textit{Proof:} See appendix \ref{proof_oftheorem2} for details.\hfill$\blacksquare $
 \subsection{Constrained Minimization Problems of Followers} \label{subsectionofgeneralized}
   \par  In this subsection, we discuss a situation that there exists inequality and equality constraints in minimization
 problems of followers. 
{Due to the constraints imposed on followers, 
$\Phi^j(x)$, $\forall j \in \mathcal{I}_L$ may become non-differentiable, posing challenges.  
 }
  \subsubsection{Equality and Inequality Constraints in Minimization Problems of Followers}
   \par  The optimal problem (\ref{followergame}) is now with constraints as 
   \begin{equation}\label{omegai}
   	\begin{aligned}
   		\Omega_i:=\{A^iy^i=b^i, 
   		h^i_r(y^i,x) \leq 0, \, r\in \mathcal{S}:= \{1,2,\dots,\tilde{s}\}      \}. 
   	\end{aligned}
   \end{equation}
   \par By the logarithmic barrier function introduced in  \cite{boyd2004convex,gould2016differentiating}, 
  we transform the optimal problem (\ref{followergame}) with $\Omega_i$ to a equivalent problem as follows:	
  \begin{equation} \label{newfollowerproblem} 
  	\left\lbrace 
  	\begin{aligned}
  		\underset{ y^i } {\operatorname{min}} \quad & 
  		s^i_R (y^i, x) := \vartheta^i s^i (y^i, x ) + \phi^i(y^i,x),   \\ 
  		s.t. \quad
  		&  A^i y^i=b^i,\, \forall i \in \mathcal{I}_F  
  	\end{aligned}
  	 	\right. 
  \end{equation}
  where $\vartheta^i \in \mathbb{R}_+$ is a scalar and 
  $\phi^i (y^i,x) =-\sum_{r=1}^{\tilde{s}} -\log(h^i_r (y^i,x))  $ with 
  $\text{dom}\phi = \{ y^i \in \mathbb{R}^{p^i} \,| \, h^i_r(y^i,x)<0, \forall r \in \mathcal{S}\}$. 
  To satisfy the strict feasibility requirements of the interior point method, we need to make the following assumptions.
  \begin{assumption} \label{assumptionofwei}
  	For all $i \in \mathcal{I}_F$, there exists positive constants $\underline{\delta}_{\vartheta^i}$ and $\overline{\delta}_{\vartheta^i}$ such that the following inequalities hold with a 
  	given positive $\vartheta^i$,  
  	\[
  	-\underline{\delta}_{\vartheta^i} \leq h^i_r( y^{i,*}_{\vartheta^i}(x) , x  ) \leq -\overline{\delta}_{\vartheta^i} <0, \forall x \in \mathbb{R}^q, \forall r \in \mathcal{S},
  	\] 
  \end{assumption} 
  where the $y^{i,*}_{\vartheta^i}(x) , x  ) $ is a 
  best response function of follower $i$ with 
  $\Omega_i$.   	
   Note that under Assumption \ref{assumptionofwei},  for every given $\vartheta^i >0$, $ s^i_R (y^i, x) $ is a strong convex function, and fulfills Assumptions  
  \ref{assumptionoffunction} and \ref{assumption_lipschitz} over $\text{dom}\phi$.
  \par By the barrier method of interior-point algorithm, we exploit the following step instead of  (\ref{algorithmline1}) for every follower $i$ as in Algorithm \ref{algorithm2}, 
  \begin{equation} \label{algorithm1.5}
  	     y^i_{\vartheta^{i},k} = \underset{ y^i \in \mathbb{R}^{p^i}, A^iy^i = b^i } {\operatorname{argmin}}
  	 \vartheta^i s^i (y^i, x ) + \phi^i(y^i,x).  
  \end{equation}
  In practice, (\ref{algorithm1.5}) is usually proceeded by an oracle, as described in Algorithm \ref{algorithm2}, which is referred to as the sequential unconstrained minimization technique (SUMT).
  In Algorithm \ref{algorithm2}, $T_g$ is the iteration counts when $  y^i_{\vartheta^{i},k}$
  is obtained with $ \vartheta^i=\vartheta^{i(g)} $. Step 1 can be solved by Gradient or Newton
  method, step 2 is a warming up step, step 3 is a terminate judgment step for a required tolerance $\epsilon^i$, and step 4 is a incremental step with a positive scalar $\chi^i$.  
  \begin{algorithm} 
  	\caption{Optimization oracle of follower $i$} \label{algorithm2}
  	for every $i\in \mathcal{I}_F$, each follower $i$ selects its strictly feasible $ y^i $, 
  	$\vartheta^i := \vartheta^{i(0)}$, $\chi^i > 1$, tolerance $\epsilon^i > 0$,
  	\par \textbf{Iteration}:
  	\qquad \par \text{Step 1}: (\ref{algorithm1.5}) with $\vartheta^i=\vartheta^{i(g)}$,  
  	\qquad \par \text{Step 2}: $ y^i_{0,\vartheta^{i(g+1)},k}  := y^i_{T_g,\vartheta^{i(g)},k}   $, 
  	\qquad \par \text{Step 3}: break if $ \vartheta^{i(g)} > \tilde{s} / \epsilon^i$,
  	\qquad \par \text{Step 4}: $\vartheta^{i(g+1)} := \chi^i \vartheta^{i(g)} $.
  \end{algorithm}
\begin{assumption} \label{assumptionof_slatercondition}
	For all $i\in \mathcal{I}_F$, $r\in \mathcal{S}$, and $\forall x \in \mathbb{R}^q$,
	 $h^i_r( y^{i} , x  )$ is convex in $y^i$ and twice continuously differentiable.
	 The matrix $A^i$ in equality constraint has complete row rank. Moreover, there exist $y^i \in \mathbb{R}^{p^i}$ 
	 such that $h^i_r ( y^{i} , x  ) <0 $, $A^i y^i = b^i $. 	
\end{assumption}
 \begin{lemma} \label{additional_dual}
	Suppose Assumptions \ref{assumptionofwei} and \ref{assumptionof_slatercondition} hold,
	then we have  $  \| y^{i,*}(x)  -   y^{i,*}_{\vartheta^i}(x)  \| \leq \sqrt{\frac{2\tilde{s}}{\mu \vartheta^i}}  $.	
	Moreover, $\|  \theta^j(x, y^*(x)) - \theta^j(x, y^*_\vartheta (x) ) \|  \leq  \ell_{\theta,0}  \sqrt{\sum_{i=1}^{N} \frac{2\tilde{s}}{\mu \vartheta^i}}  $ holds.
\end{lemma}
\par \textit{Proof:} See appendix \ref{proof_ofadditional_dual} for details.\hfill$\blacksquare $
\par Note that with the increasing of each $\vartheta^i $, $i\in \mathcal{I}_F$, the gap is decreasing. 
When $\vartheta^i  \rightarrow \infty $, $i\in \mathcal{I}_F$, then $  y^*(x) \rightarrow   y^*_\vartheta (x)$.
\begin{proposition} \label{implict_gradient}
	Suppose Assumptions \ref{assumptionofwei} and \ref{assumptionof_slatercondition} hold, with each follower's minimization problem as 
	(\ref{newfollowerproblem}),   
	$ J^{j,*}_{\mathcal{P}_h,k} $ and $H^{*}_{\mathcal{P}_h,k} $ defined in Lemma \ref{lemma2} are reformulated as 
	$ J^{j,*}_{\mathcal{P}_h,k} := \text{diag}(\nabla_{x^j}\nabla_{z}s^i_R(z ,x_k)_{i\in \mathcal{P}_h})|_{z=y^{i,*}(x_k)}$ and
	$ H^{*}_{\mathcal{P}_h,k} := \text{diag}((\tilde{H}^{i,*}_{R})_{i\in \mathcal{P}_h})   $ respectively, 
	where $ \tilde{H}^{i,*}_{R} = ({H^{i,*}_{R}})^{-1}-({H^{i,*}_{R}})^{-1} ({A^i})^\top [ {A^i} ({H^{i,*}_{R}})^{-1} ({A^i})^\top ]^{-1} {A^i} ({H^{i,*}_{R}})^{-1}$,
	$ H^{i,*}_{R} = \nabla^2_{y^i} s^i_R(y^i,x_k)|_{y^i=y^{i,*}(x_k)}$. 
	Moreover, we have 
     $\nabla_{x^j}\nabla_{z}s^i_R(z ,x_k)|_{z=y^{i,*}(x_k)} = \vartheta^i \nabla_{x^j}\nabla_{z} s^i(z,x_k) - \nabla_{x^j}\nabla_{z}\phi^i(z,x_k)|_{z=y^{i,*}(x_k)}$  
     and $ \nabla^2_{y^i}s^i_R(y^i ,x_k)|_{y^i=y^{i,*}(x_k)} = \vartheta^i \nabla^2_{y^i} s^i(y^i,x_k) - \nabla^2_{y^i}\phi^i(y^i,x_k)|_{y^i=y^{i,*}(x_k)}$.
\end{proposition}
\par \textit{Proof:} See appendix \ref{proof_ofimplict_gradient} for details.\hfill$\blacksquare $
\par Note that by Lemma \ref{implict_gradient}, the SE of the MLMF Stackelberg game with 
followers' minimization problems as (\ref{newfollowerproblem}) can be  solved by Algorithm \ref{algorithm1} and Oracle \ref{algorithm2}. To simplify, let $\vartheta^i=\vartheta^l=\vartheta, \forall i,l \in \mathcal{I}_F$.
  \subsubsection{Example of Rectangle constraints }
   \par We discuss a specialized convex set, rectangle constraints, in followers' minimization problems, i.e, $ [y^i]_r - [u^i]_r \leq 0, -[y^i]_r + [l^i]_r \leq 0, \, r\in \{ 1,2,\dots, p^i \}$,
   where $[y^i]_r$ is the $r$-th item of $y^i$. Moreover, $l^i$ and $u^i$ are constant vectors that are the lower and upper bound of $y^i$. 
   Thus, (\ref{newfollowerproblem}) is now reformulated as 
   \begin{equation}
   \begin{aligned}
   		   	\underset{ y^i } {\operatorname{min}}  \quad& 
   		s^i_R (y^i, x) := \vartheta^i s^i (y^i, x ) + \phi(y^i),   \\ 
   \end{aligned}
   \end{equation}
   where $\vartheta^i \in \mathbb{R}_+$ is a scalar and 
   $\phi (y^i) =-\sum_{r=1}^{p^i} \log ([y^i]_r - [l^i]_r )  - \sum_{r=1}^{p^i} \log ([u^i]_r - [y^i]_r)  $ with 
   $\text{dom} \phi = \{ y^i \in \mathbb{R}^{p^i} \, | \,   l_i  <  y^i <  u^i \}$.
   Note that for every given $\vartheta^i >0$, $ s^i_R (y^i, x) $ is a strong convex function, and fulfills Assumptions  
   \ref{assumptionoffunction} and \ref{assumption_lipschitz} over $\text{dom}\phi$.
   Moreover, the gradient decent method can also be utilized in Algorithm \ref{algorithm2}, 
   \begin{equation} \label{algorithm1.75}
  y^i_{t+1,\vartheta^{i},k} = 
  y^i_{t,\vartheta^{i},k} - \alpha_R \nabla_{y^i} s^i_R (y^i_{t,\vartheta^{i},k},x_k ), 
   \end{equation}
   where $ \alpha_R $ is set to be the same as Lemma \ref{lemmaofyTtoyx} with the function to be 
   $s^i_R (y^i, x)$ instead of $s^i (y^i, x)$. 
 \section{Numerical Examples}  \label{section5}
 \par In this section, we present 
  numerical simulations of a microgrid management system and a class of heterogeneous cellular networked system.
\subsection{Case 1: Networked Stackelberg-Cournot Equilibrium Problems}
\par To verify the effectiveness of Algorithm \ref{algorithm1}, numerical experiments are conducted on a system consisting of 4 microgrids, with each corresponding cluster consisting of 5, 6, 7 and 8 users, respectively.  The topology of microgrids is presented as
\begin{equation}
	 4 \rightleftharpoons 1 \rightleftharpoons 2 \rightleftharpoons 3, \quad 2 \rightleftharpoons 4. \notag
\end{equation} 
\par
\begin{figure}[htbp]
	\centering
	\begin{minipage}{0.49\linewidth}
		\centering
		\includegraphics[width=1.1\linewidth]{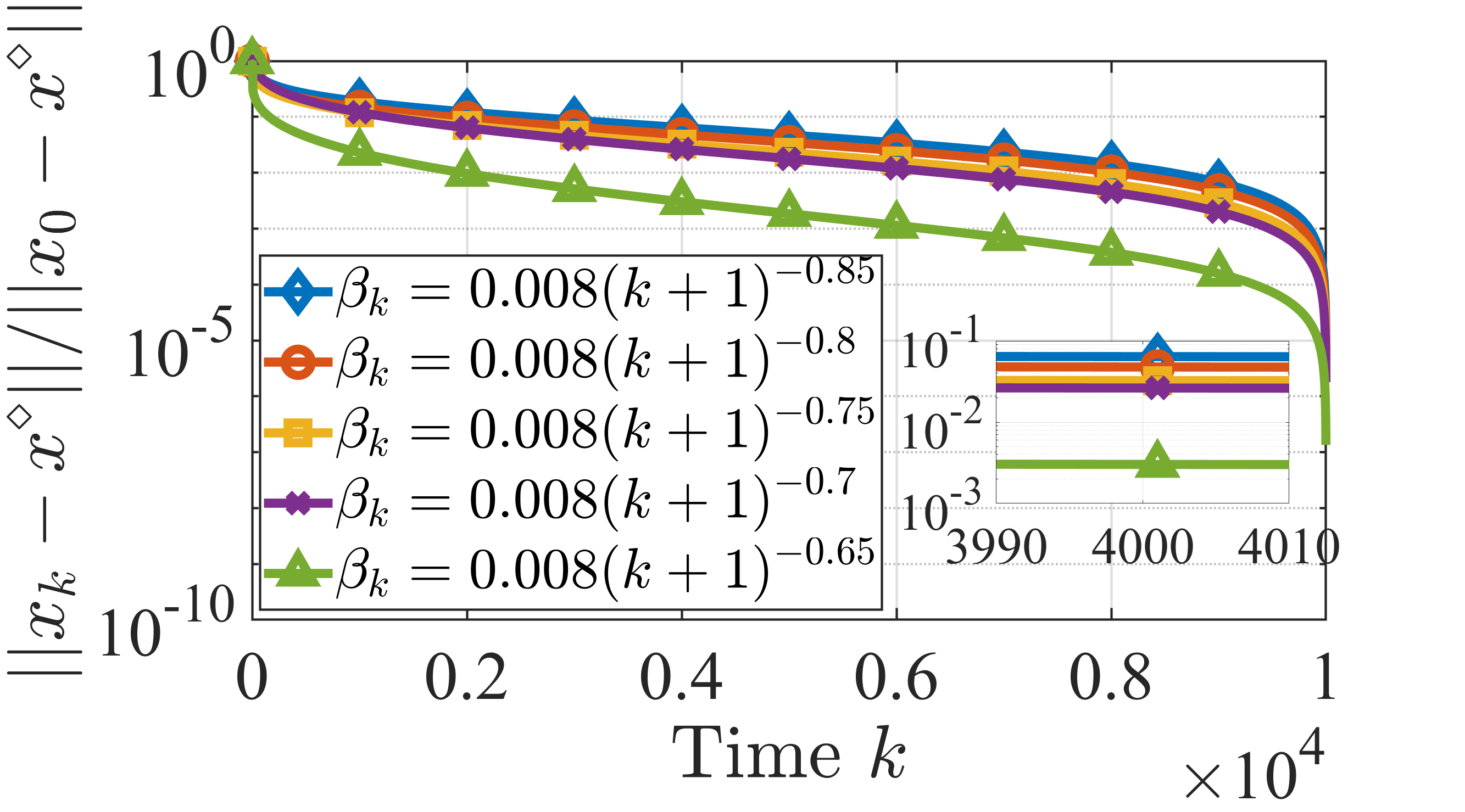}
	\end{minipage}
	\begin{minipage}{0.49\linewidth}
		\centering
		\includegraphics[width=1.1 \linewidth]{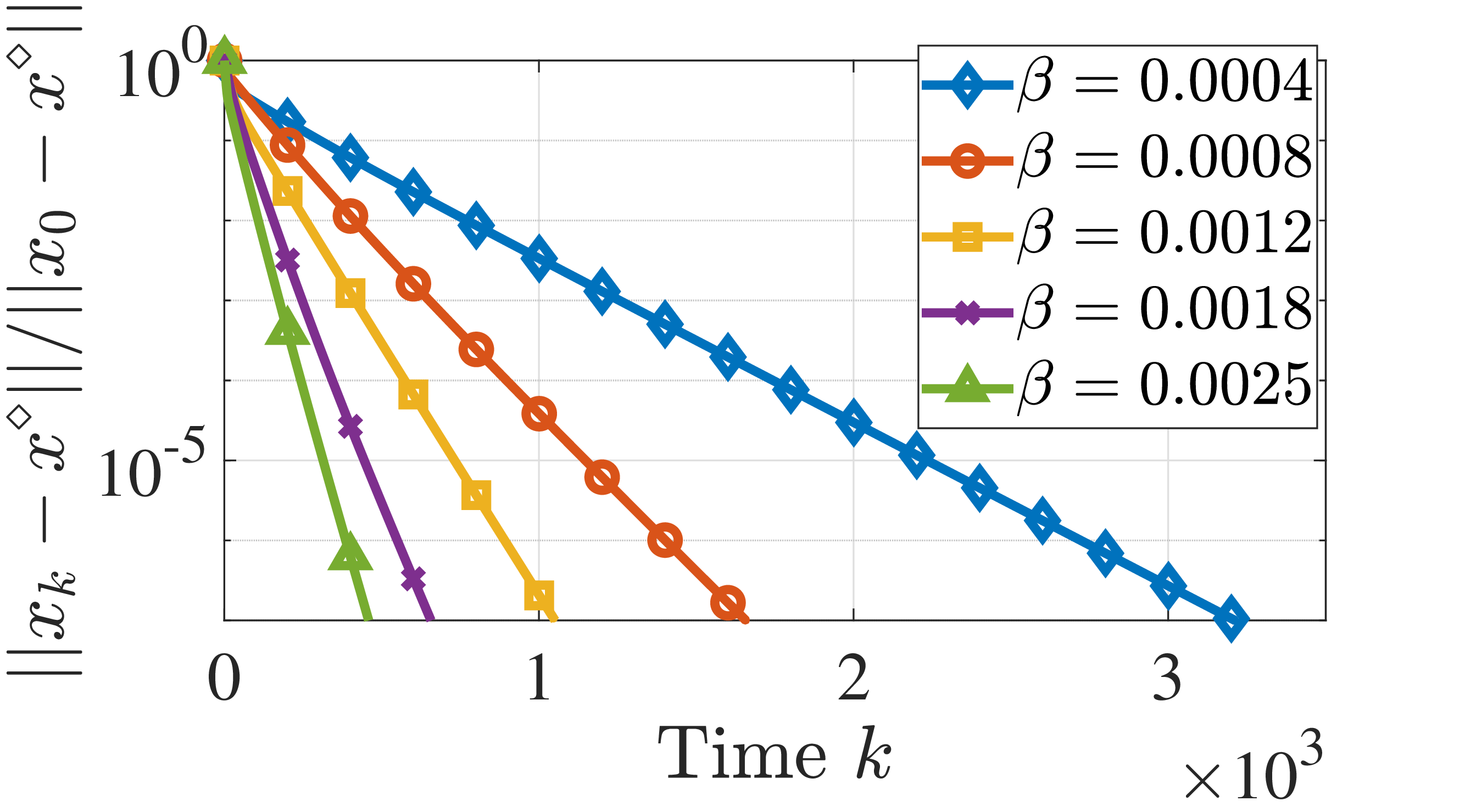}
	\end{minipage}
	\caption{The  trajectories of 
		$ \frac{\| x_k -  x^\diamond   \|}{ \|x_0 - x^\diamond\|} $ with different diminishing steps $\beta_k$ (left) and constant steps $\beta$ (right).} \label{convergent1}
\end{figure}

 
The values for $T$, $D$, and $B$ are set to be $6000$, $6000$, and $8000$, respectively, to satisfy Proposition \ref{lemmaofx0z0zhat0}.
It can be verified that this case fulfills both diminishing and constant step-size scenarios.
As depicted in the left side of Fig.\ref{convergent1},  $ x_k $ generated by Algorithm \ref{algorithm1} converges under various diminishing step-sizes $\beta_k$ with different speeds. Analogously, $ x_k $ converges for different values of $\beta$. It can be inferred that, under Assumption \ref{assumptionofstronglymonotone} and $ \Omega = \mathbb{R}^q $, the sequence $ x_k $ converges with different constant step sizes at a linear rate, as illustrated in the right side of Fig.\ref{convergent1}.  
\par
\begin{figure}[htbp]
	\centering
	\begin{minipage}{0.49\linewidth}
		\centering
		\includegraphics[width=1.1\linewidth]{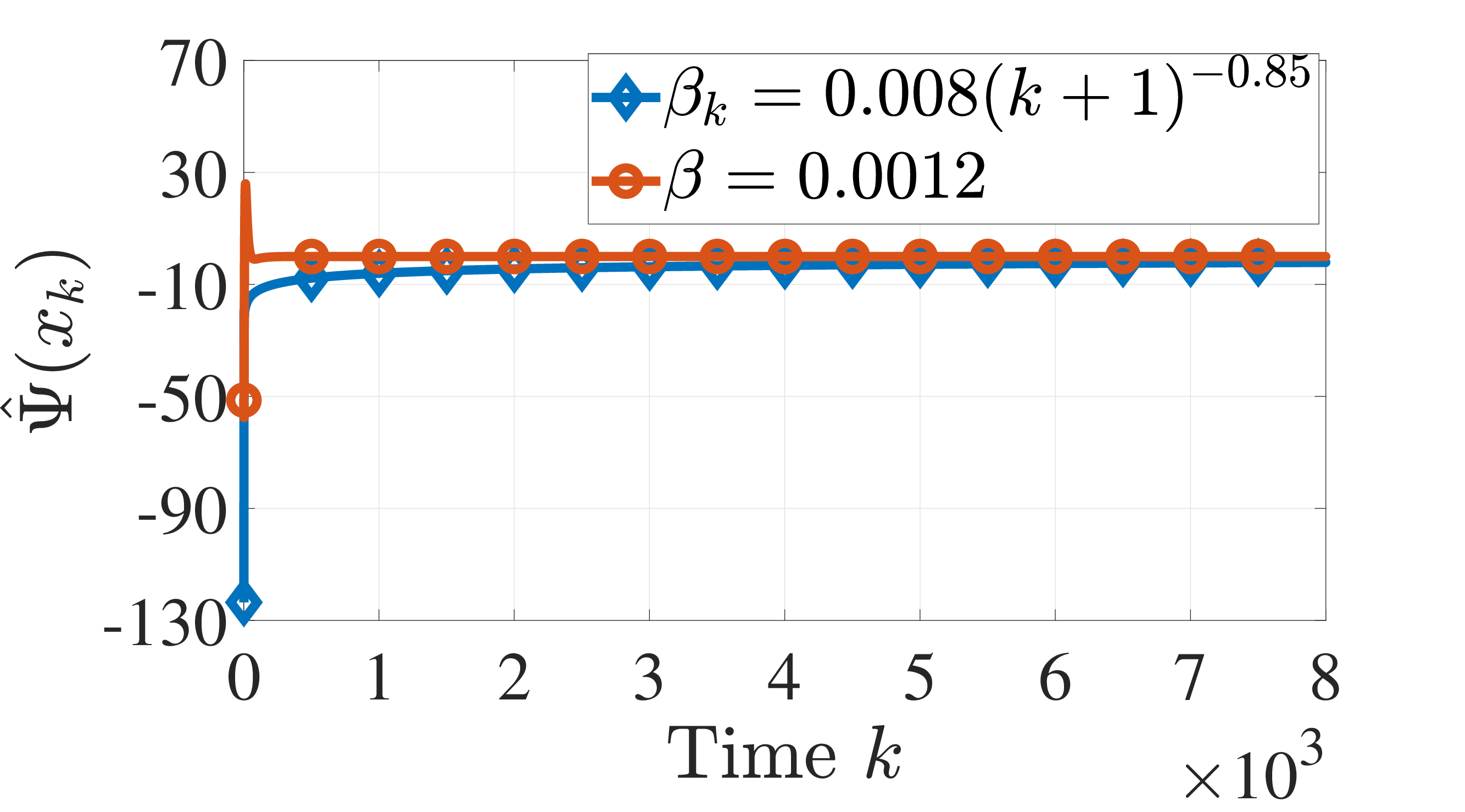}
	\end{minipage}
	\begin{minipage}{0.49\linewidth}
		\centering
		\includegraphics[width=1.1 \linewidth]{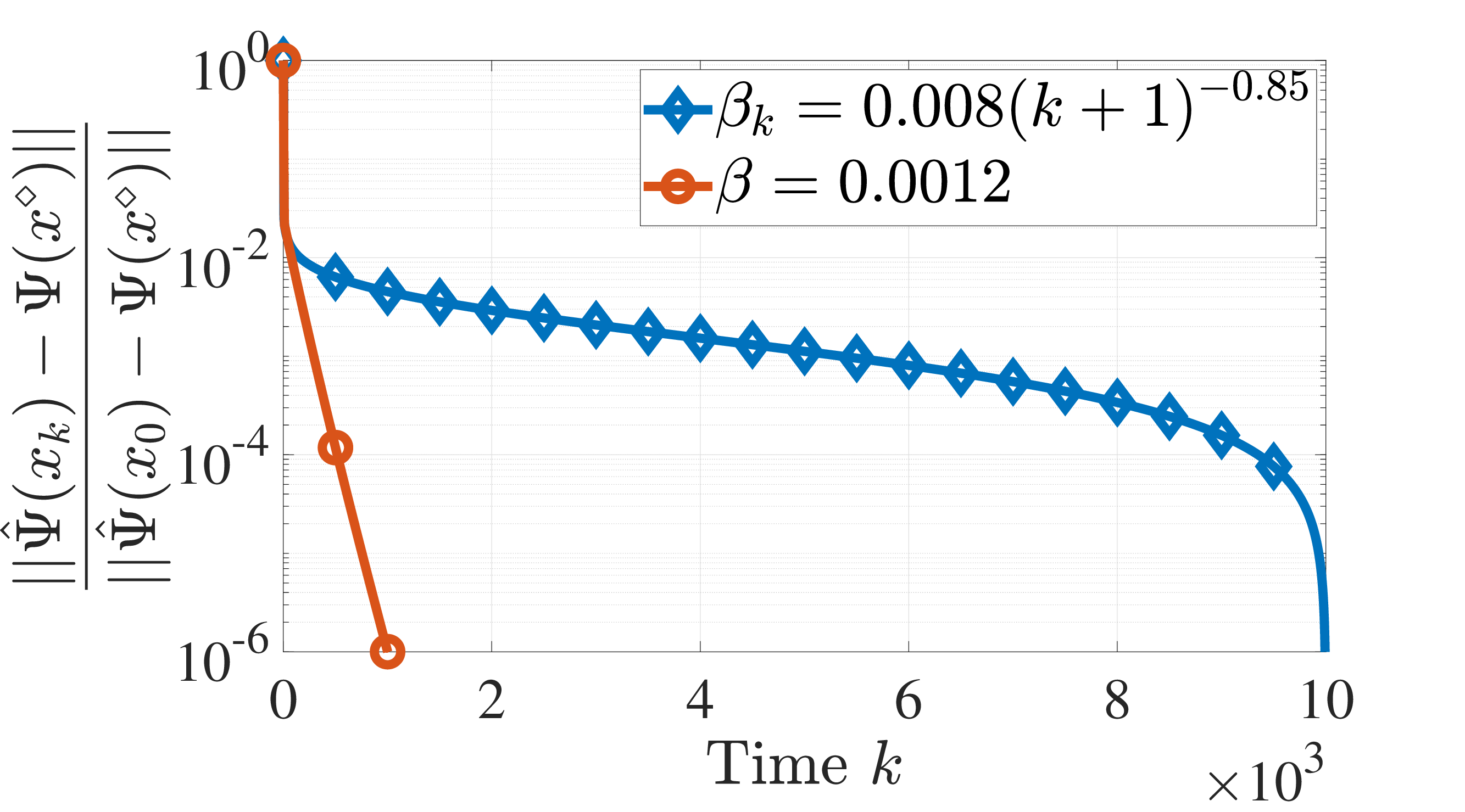}
	\end{minipage}
		\caption{The  trajectories of 
			$ \hat{\Psi}(x_k) $ (left)  and 
			$ \frac{\|\hat{\Psi}(x_k)  -  \Psi(x^\diamond)  \|}{ \|\hat{\Psi}(x_0)  - \Psi(x^\diamond) \|}    $ (right)
			with diminishing step $\beta_k$ and constant step $\beta$.}
	\label{hatpsi}
\end{figure}

 Moreover, as depicted in Fig.\ref{hatpsi}, $\Psi(x_k)$ converges more fast with a constant step-size than with a diminishing step-size, even though the initial value $\beta_0\approx0.0044$ is greater than the constant step-size $0.0012$.
\subsection{Case 2: Networked Stackelberg-Cournot Equilibrium Problems}
Without loss of generality, we set the dimension of follower $i \in \mathcal{I}_F$ to $1$ for simplicity. 
Analogously, let the number of operators be 4, and each operator has a cluster with 5, 6, 7, and 8 users.
The values for $T$, $D$, and $B$ are also configured as $6000$, $6000$, and $8000$, and the topology of operators' network is set to be the same as Case 1. 
\par
\begin{figure}[htbp]
	\centering
	\begin{minipage}{0.49\linewidth}
		\centering
		\includegraphics[width=1.1\linewidth]{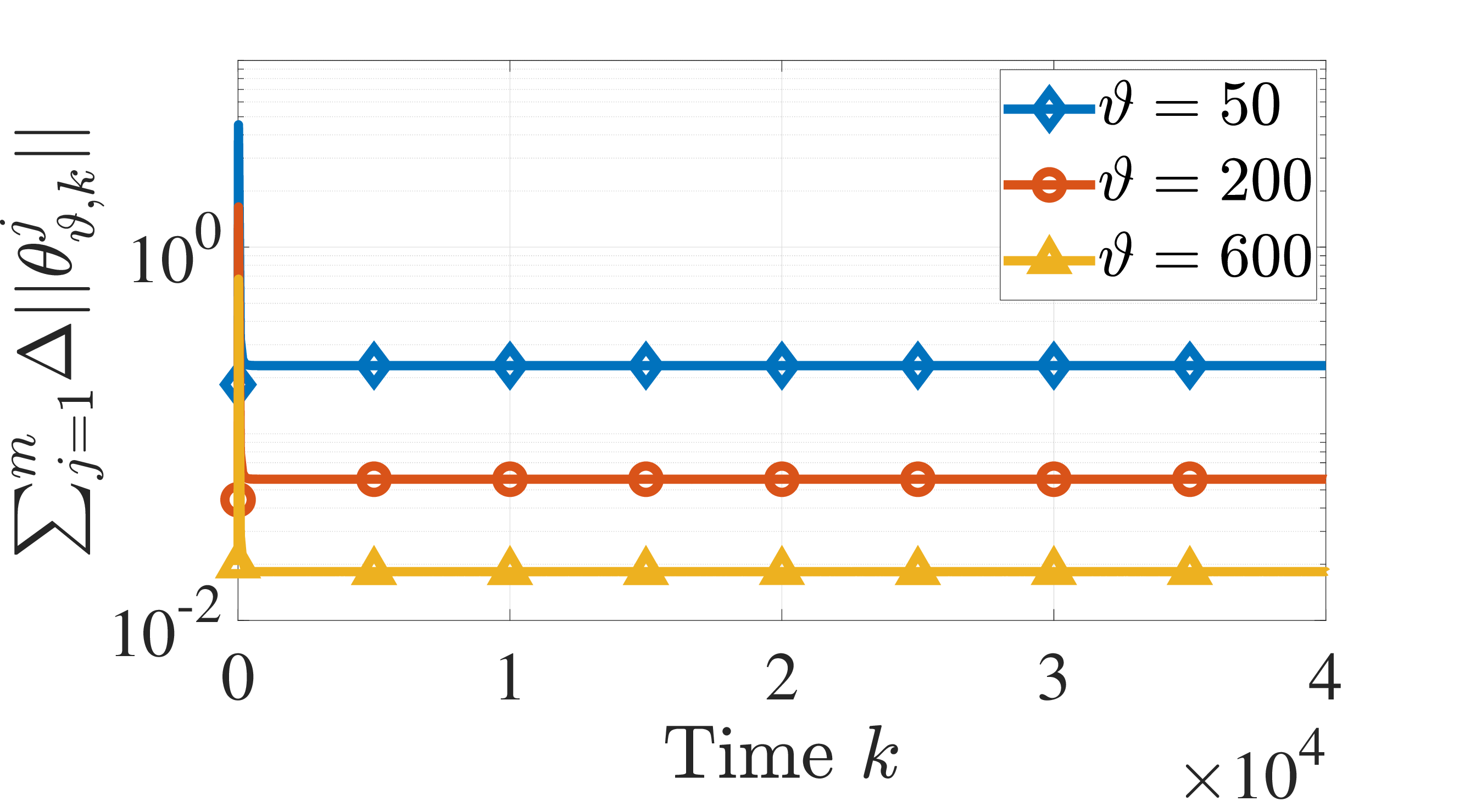}
	\end{minipage}
	\begin{minipage}{0.49\linewidth}
		\centering
		\includegraphics[width=1.1 \linewidth]{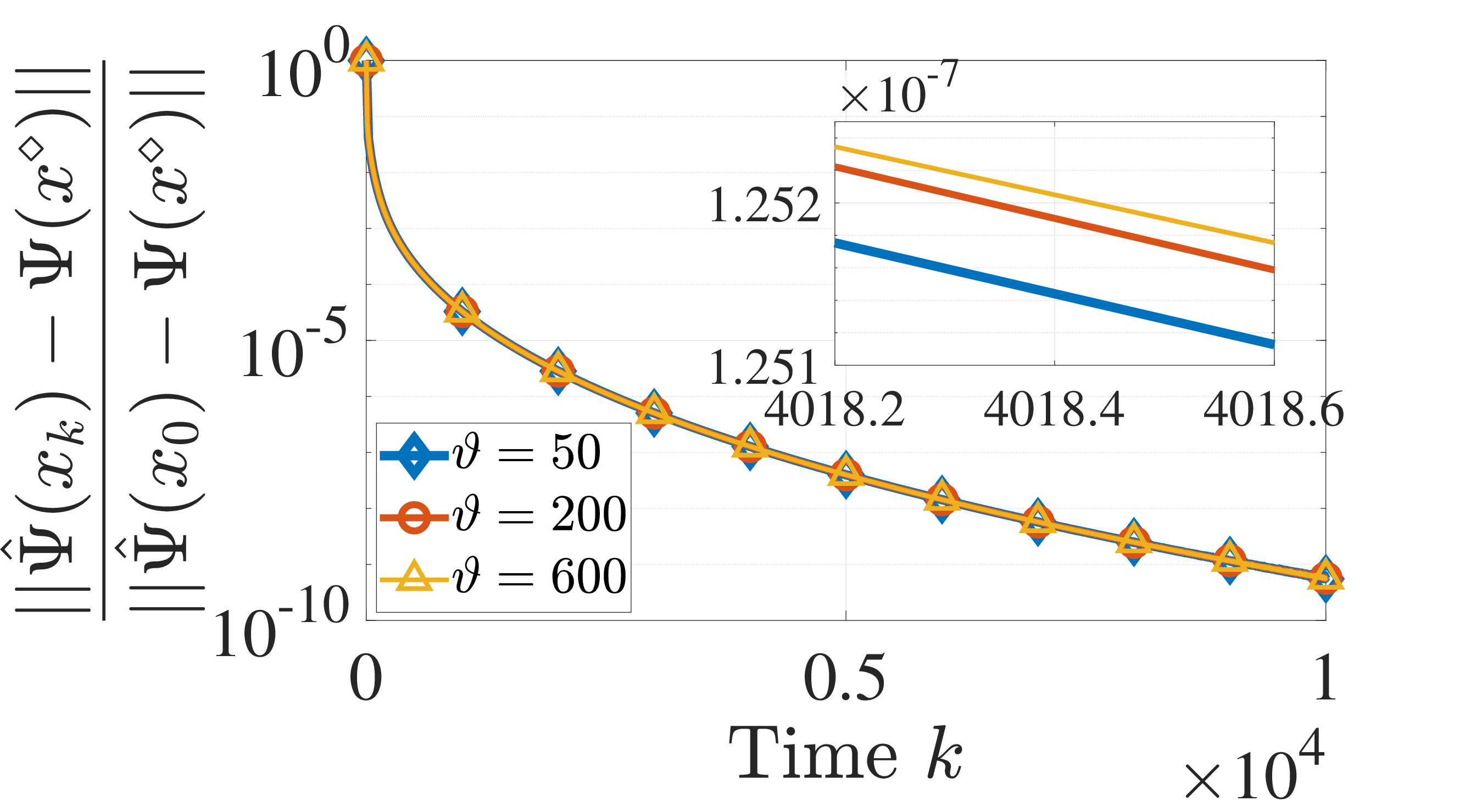}
	\end{minipage}
		\caption{The  trajectories of 
			$ \textstyle{\sum_{j=1}^{m}}\Delta\| \theta^j_{\vartheta,k} \| $ (left) and   
			$ \frac{\|\hat{\Psi}(x_k)  -  \Psi(x^\diamond)  \|}{ \|\hat{\Psi}(x_0)  - \Psi(x^\diamond) \|}    $ (right)   with different values of $\vartheta$.} \label{comm}
\end{figure}

 We define $ \textstyle{\sum_{j=1}^{m}}\Delta\| \theta^j_{\vartheta,k} \| : =\textstyle{\sum_{j=1}^{m}}\| \theta^j(x_k,y^*_\vartheta (x_k)   )  -  \theta^j(x_k,y^*(x_k)   )\|   $. As depicted in the left side of Fig.\ref{comm}, 
a larger $\vartheta$ results in a smaller $ \textstyle{\sum_{j=1}^{m}}\Delta\| \theta^j_{\vartheta,k} \|$, which verifies 
Lemma \ref{additional_dual}. 
Moreover, the right side of Fig.\ref{comm} 
shows that a smaller $\vartheta$ yields a smaller error $ \frac{\|\hat{\Psi}(x_k)  -  \Psi(x^\diamond)  \|}{ \|\hat{\Psi}(x_0)  - \Psi(x^\diamond) \|}$, which implies the
trade off relationship between the larger $\vartheta$ and the worse convergence performance.
\section{Conclusion} \label{section6}
\par 
In this paper, we have developed a distributed SE seeking algorithm for networked MLMF Stackelberg games. 
Motivated by the need for the implementation of Stackelberg games, 
the information structures of networked systems have to be taken into consideration.
Due to the increasing number of players and expanding scenarios,
a clustered information structure  arises and imposes difficulties
for seeking the SE distributively.
In conjunction with implicit gradient estimation and network consensus mechanisms, an algorithm for seeking SE distributively for MLMF games is proposed.
We have rigorously proven the convergence of our algorithm with both diminishing and constant step sizes under strict and strong monotonicity assumptions when followers' cost functions are strongly convex. 
Furthermore, we have explored the SE seeking approach when the minimization problems of followers have linear equality and inequality constraints.
At last, we have conducted numerical simulations to validate our proposed framework. 
Overall, our algorithm provides an effective solution to address the challenges posed by networked systems modeled by MLMF Stackelberg games with large populations.
\appendix
 
\section{Appendix}  
\subsection{ Proof of \textbf{Lemma \ref{lemma2}}} \label{proofoflemma2} 
\par Based on the optimality of $ i $-th follower, we have  $ \nabla _ {y^i} s^i (y^i, x_k)|_{y^i=y^{i,*}(x_k)} =0$.  Afterwards, we use the implicit differentiation w.r.t. $ x^j  $, and it yields 
\small
\begin{flalign}
	\nabla_{x^j} \nabla_{z}s^i(z, x_k ) 
	+  \frac{\partial z}{\partial x^j } \nabla^2_{z} s^i (z, x_k )|_{z=y^{i,*}(x_k)} 
	=0. \notag
\end{flalign}
\normalsize
It followers that 
\small
\begin{flalign} \label{temp1}
	\frac{\partial z}{\partial x^j }=-\nabla_{x^j} \nabla_{z}s^i(z, x_k ) (\nabla^2_{z} s^i (z, x_k ))^{-1}|_{z=y^{i,*}(x_k)},
\end{flalign}
\normalsize
where the equality holds in light of the $ \mu  $-strong convexity of $ s^i(y^i,x) $, which implies that
$ \nabla^2_{y^i} s^i (y^{i}, x ) \succ \mu I$ .
We sequentially stack both sides of (\ref{temp1}) where $ i\in \mathcal{P}_h $ to form a diagonal block matrix, which results in 
\small
\begin{flalign} \label{temp2}
	\text{diag}
	\left( 	
	\left(\frac{\partial y^{i,*}(x)}{\partial x^j }\right)_{i\in \mathcal{P}_h}	\right)	
	= - J^{j,*}_{\mathcal{P}_h,k} (H^{*}_{\mathcal{P}_h,k})^{-1},
\end{flalign} 
\normalsize
where the equality holds in light of the inverse properties of the diagonal block matrix. 
Furthermore, we stack both sides of (\ref{temp2}) from $ j=1 $ to $ m $, which further implies the equality (\ref{lemma1}) holds.\hfill $ \blacksquare $
\subsection{ Proof of \textbf{Lemma \ref{lemmaofhypergradient}}} \label{proofoftheorem2} 
Foremost, we have
\small
\begin{flalign}
	J^{j,*}_{\mathcal{P}_h,k}	 (H^{*}_{\mathcal{P}_h,k})^{-1}\notag 
	=R^j_h J^*_{\mathcal{P}_h,k} (\boldsymbol{I}_m \otimes H^*_{\mathcal{P}_h,k})^{-1}  U^j_h, 
\end{flalign}
\normalsize
and stack up all items from  $ h=1  $ to $ m $ on the diagonal of a matrix.
It yields
\small
\begin{flalign}
	\text{diag}(  ( J^{j,*}_{\mathcal{P}_h,k}  )_{h\in \mathcal{I}_L})
	\text{diag}(  (   H^{*}_{\mathcal{P}_h,k}  )_{h\in \mathcal{I}_L}  )^{-1} = R^j J^*_k (H^*_k )^{-1} U^j. \notag
\end{flalign} 
\normalsize
In light of Lemma 2.2 in \cite{ghadimi2018approximation}, we have the Lipschitz continuous property of $y^{*}(x) $, i.e., $\|	y^{*}(x) - y^{*}(\bar{x})  \| \leq  \sqrt{N}\kappa \| x- \bar{x} \|$, 	where $ \kappa=\ell /\mu  $.
Furthermore, we invoke the definition of hyper-gradient (\ref{definitionofhypergradient}), and in conjunction with 
the Lipschitz continuous property of $ y^*(x) $ to form the following equality:
\small
\begin{flalign}
	&\nabla_{x^j} \Phi^j(x):=\nabla_{x^j} \theta^j(x, y)  
	-E^j R^j J^*_k (H^*_k )^{-1} U^j 
	\nabla_y \theta^j (x, y)|_{y=y^*(x)}.  \notag
\end{flalign}
\normalsize
Additionally, it follows that   (\ref{lemmaline1})  is hold by stacking $ \nabla_{x^j} \Phi^j(x) $ 
form $ j=1 $ to $ m $ on the diagonal of a matrix.
The proof is finished. \hfill $ \blacksquare $

\subsection{Proof of \textbf{Lemma \ref{lemmaofinverse}}} \label{proofoflemmaofinverse}
We have 
\small
\begin{equation}
	\begin{split}
		&	\|    Z_{\mathcal{P}_j,d,k}   - Z^{R}_{\mathcal{P}_j,k} \|_F  \leq 
		\|    Z_{\mathcal{P}_j,d-1,k}   -  \gamma  Z_{\mathcal{P}_j,d-1,k}  (\boldsymbol{I}_m \otimes H_{\mathcal{P}_j,k}) +\notag \\
		& 	   \gamma J_{\mathcal{P}_j,k} - Z^{R}_{\mathcal{P}_j,k}  \|_F  \leq \| ( Z_{\mathcal{P}_j,d-1,k}    - Z^{R}_{\mathcal{P}_j,k} ) 
		(I-\gamma (\boldsymbol{I}_m \otimes H_{\mathcal{P}_j,k})) \\
		&- \gamma Z^{R}_{\mathcal{P}_j,k} (\boldsymbol{I}_m \otimes H_{\mathcal{P}_j,k}) + \gamma J_{\mathcal{P}_j,k} \| \\
		&\leq^{(a)} \|   Z_{\mathcal{P}_j,d-1,k}    - Z^{R}_{\mathcal{P}_j,k}  \|_F 
		\| I-\gamma (\boldsymbol{I}_m \otimes H_{\mathcal{P}_j,k})   \|_F
	\end{split}
\end{equation} 
\normalsize
where $ (a) $ holds in light of $ Z^{R}_{\mathcal{P}_j,k}:= J_{\mathcal{P}_j,k} (\boldsymbol{I}_m \otimes H_{\mathcal{P}_j,k})^{-1} $, and in conjunction with the Cauchy-Schwarz inequality.
Since 
$ \| I-\gamma (\boldsymbol{I}_m \otimes H_{\mathcal{P}_j,k})   \|_F \leq^b   \sqrt {m \sum_{j=1}^{p_{\mathcal{P}_j}} ( 1-\gamma \delta_j  )^2  } \leq \sqrt {m p_{\mathcal{P}_j} (1-\gamma \mu)^2  }$, where $b  $ holds in light of $ \frac{1}{\mu}-\frac{1}{\sqrt{m p_M}\mu }    <\gamma \leq  \frac{1}{\ell_{ s,2 }} $.
Thus, the proof is finished. \hfill $ \blacksquare $

\subsection{Proof of \textbf{Lemma \ref{lemmaofZ}}} \label{proofoflemma10}
By (\ref{algorithmestimate}), we have 
\small
\begin{equation}
	\begin{split}
		&\hat{Z}^{j(h)}_{b,k} = (\sum _{g\in \mathcal{N}_j} w^{jg} \hat{Z}^{g(h)}_{b-1,k} )  + \xi^j_h w^{jh} ( Z_{\mathcal{P}_h,D,k} -  \hat{Z}^{j(h)}_{b-1,k}  )   \notag\\
		&- \sum _{g\in \mathcal{N}_j}w^{jg} Z_{\mathcal{P}_h,D,k}  +  Z_{\mathcal{P}_h,D,k} =
		( \sum _{g\in \mathcal{N}_j} w^{jg} \hat{Z}^{g(h)}_{b-1,k}   -  \\
		& \sum _{g\in \mathcal{N}_j} w^{jg}  Z_{\mathcal{P}_h,D,k}  )  +  \xi^j_h w^{jh} ( Z_{\mathcal{P}_h,D,k} -  \hat{Z}^{j(h)}_{b-1,k}  ) +  Z_{\mathcal{P}_h,D,k}.
	\end{split}
\end{equation}
\normalsize
In light of the definition of $\tilde{W}^h$, the above equality is equivalent to 
\begin{equation}
	\hat{Z}^{j(h)}_{b,k}  = \sum _{g=1}^m \tilde{w}^{jg}_h ( \hat{Z}^{g(h)}_{b-1,k}  -   Z_{\mathcal{P}_h,D,k}   ) + Z_{\mathcal{P}_h,D,k},
\end{equation}
Moreover, we stack up all leaders' estimates to a specific leader, i.e., $ h \in \mathcal{I}_L $, to yield the following equality 
\begin{equation}
	\begin{split}
		\hat{Z}^{:(h)}_{b,k}  - \boldsymbol{1}_m \otimes Z_{\mathcal{P}_h, D,k } = \tilde{\boldsymbol{W}} (\hat{Z}^{:(h)}_{b-1,k}  - \boldsymbol{1}_m \otimes Z_{\mathcal{P}_h, D,k }). \notag
	\end{split}
\end{equation}
Furthermore, we stack up the above equality from $ h=1 $ to $ m $ to yield 
\begin{equation}
	\hat{Z}^{:(:)}_{b,k} - Z_{:,D,k} = \tilde{\boldsymbol{W}} (  \hat{Z}^{:(:)}_{b-1,k}   -  Z_{:,D,k}    ),
\end{equation}
and we take the $ \omega $-norm on both sides 
\begin{equation}
	\begin{split}
		&\| \hat{Z}^{:(:)}_{b,k} - Z_{:,D,k}  \|_w  = \|  \tilde{\boldsymbol{W}} (  \hat{Z}^{:(:)}_{b-1,k}   -  Z_{:,D,k}    )  \|_w \\
		&\leq^{(a)} \| \tilde{\boldsymbol{W}}  \|_w   \|  \hat{Z}^{:(:)}_{b-1,k}   -  Z_{:,D,k}   \|_w,
	\end{split}
\end{equation}
where $ (a) $ holds in light of the Cauchy-Schwarz inequality and the compatibility principle of matrix norms.
Furthermore, in conjunction with Th.5.6.10 in \cite{horn2012matrix},
it can be verified that there alway exist a norm, i.e., $ \omega $-norm of $ \tilde{\boldsymbol{W}} $, having the value 
$ \sigma_w:= \| \tilde{\boldsymbol{W}}   \|_w  < 1 $, since the spectrum radius $ \rho(\tilde{\boldsymbol{W}}) <1 $ holds and  $ \rho(\tilde{\boldsymbol{W}}) \leq  \| \tilde{\boldsymbol{W}}   \|_w \leq   \rho(\tilde{\boldsymbol{W}}) + \epsilon $ holds by setting small enough $ \epsilon$. Thus, the proof is finished. \hfill $ \blacksquare $

\subsection { Specific forms from $\Xi_1$ to $\Xi_8$   } \label{constantnotations}
\small
\begin{equation}
	\begin{aligned}
		\Xi_1 &:= \frac{6m^2p^2N\ell^2_{s,1} \ell^2_{s,2}}{\mu^4} +\frac{6mp\ell^2_{s,2} }{\mu^2},\,\, \Xi_2:= \Xi_1 N \kappa^2 + \Xi_1 N , \notag\\   
		\Xi_3&:=  3m\ell^2_{\theta,1} + \frac{3m^3N^2p\ell^2_{s,1}\ell^2_{\theta,1}}{\mu^2},\,\,\Xi_4 := 3N^2\ell^2_{\theta,0}\Xi_2 + \Xi_3,\notag\\
		\Xi_5&:=\Xi_3+3N^2\ell^2_{\theta,0}\Xi_1,\,\, \Xi_6 := 2m\ell^2_{\theta,0} + \frac{2m^2N^3p\ell^2_{s,1}\ell^2_{\theta,0}}{\mu^2}, \notag \\
		\Xi_7&:= \frac{16\Xi_1 N \kappa^2}{3m} + \frac{2\Xi_1 N}{3m},\,\, \Xi_8:= 2N\kappa^2 + \Xi_7  +8m\Xi_7. \notag\\
	\end{aligned}
\end{equation}
\normalsize
\subsection{Proof of \textbf{Lemma \ref{lemmaofdistanceofz}}}\label{proofoflemmaofdistanceofz}
\par Foremost, we have
	\small
	\begin{equation}
		\begin{split}
			&\| \hat{Z}_{B,k}  -  \boldsymbol{I}_m \otimes Z^*(x^{\diamond})    \|^2_F= \| \hat{Z}_{B,k}  - \boldsymbol{I}_m \otimes Z_{D,k} + \boldsymbol{I}_m \otimes Z_{D,k}  \notag   \\
			&    -\boldsymbol{I}_m\otimes Z^R_k + \boldsymbol{I}_m\otimes Z^R_k   - \boldsymbol{I}_m \otimes \bar{Z}_k  +  \boldsymbol{I}_m \otimes \bar{Z}_k  - \boldsymbol{I}_m \otimes Z^*_k  \\
			&+ \boldsymbol{I}_m \otimes Z^*_k - \boldsymbol{I}_m \otimes \tilde{Z} + \boldsymbol{I}_m \otimes \tilde{Z}-\boldsymbol{I}_m \otimes Z^*(x^{\diamond}) \|^2_F \\
			&\leq^{(a)} 6 r_z \sigma_2^{2B}  \|\hat{Z}_{0,k}   - \boldsymbol{I}_m \otimes   Z_{D,k}   \|^2_F + 6m C_D \| Z_{0,k}  - Z^R_k      \|^2_F\\
			& + 6m\|J_k \|^2_F \| H^{-1}_{k} - (H^{*}_k)^{-1}\|^2_F + 6m\| (H^{*}_k)^{-1}\|^2_F  \| J^*_k - J_k  \|_F^2 \\
			&+ 6m \| J^*_k\|^2_F \| (H^{*}_k)^{-1}  -  ( H^*( x^\diamond ))^{-1}     \|^2_F\\
			& + 6m\| ( (H^*( x^\diamond ))^{-1}\|^2_F \|   J^*_k  -  J^*(x^{\diamond})  \|^2_F,
		\end{split}
	\end{equation}
\normalsize
where 
$ \bar{Z}_k : = J_k H_k^{-1} $, 
$ \tilde{Z} :=J^*_k H^*(x^\diamond)^{-1} $,
$ Z^*(x^{\diamond}):= J^*(x^\diamond) (  H^*( x^\diamond ) )^{-1} $, 
and $ (a)  $ holds in light of the extension of $ \| a+b  \|^2 \leq  2\|a\|^2+2\|b\|^2 $.  
Moreover, in light of Assumption \ref{assumptionoffunction} and \ref{assumption_lipschitz}, 
we have $ \| J_k \|^2_F, \| J_k^* \|^2_F \leq mN\ell^2_{s,1} $,
$ \| H^{-1}_k \|^2_F, \|H^*( x^\diamond )^{-1}\|^2_F \leq  \frac{p}{\mu^2} $, 
$ \|  H^{-1}_{k} - (H^{*}_k)^{-1} \|^2_F \leq \|H^{-1}_{k}\|^2_F \|(H^{*}_k)^{-1}\|^2_F \|H_{k}-H^{*}_k\|^2_F 
\leq  \frac{\ell_{ s,2 }^2 p^2}{\mu^4}  \| y_{T,k}- y^*(x_k) \|^2  $, 
and $ \| J^*_k - J_k   \|_F \leq \ell^2_{s,2} \| y_{T,k} - y^* (x_k) \|^2 $.
Thus, we have 
\small
\begin{equation}
	\begin{split}
	&\| \hat{Z}_{B,k}  -  \boldsymbol{I}_m \otimes Z^*(x^{\diamond})    \|^2_F  \leq  	\Xi_2 \| x_k -x^\diamond  \|^2 + 6mC_D \| Z_{0,k}  - Z^R_k      \|^2_F \\
	&+\Xi_1 C_T  \| y_{0,k}- y^*(x_k) \|^2  + 6C_B \|\hat{Z}_{0,k}   - \boldsymbol{I}_m \otimes   Z_{D,k}   \|^2_F.    \notag 
	\end{split}
\end{equation}	
Let $ \hat{\Psi}(x): =  \text{diag} (  ( \nabla_{x^j} \hat{\Phi}^j(x) )_{j\in \mathcal{I}_L}  ) $, then we have
\small
\begin{equation}
\begin{split}
	&\| \hat{\Psi}(x_{k}) -  \Psi(x^\diamond)    \|^2   \leq \|\Theta_x(x_{k},y_{T,k}) -  \Theta_x(x^\diamond,y^*(x^\diamond))   \notag \\
	& -ER\hat{Z}_{B,k} U \Theta_y(x_{k},y_{T,k})+ER(\boldsymbol{I}_m \otimes Z^*(x^\diamond))U\Theta_y (x_{k},y_{T,k})\\
	&-ER(\boldsymbol{I}_m\otimes Z^*(x^\diamond))U
	(\Theta_y(x_{k},y_{T,k})-\Theta_y(x^\diamond,y^*(x^\diamond)))\|\\
	&\leq \Xi_3( \|x_{k}-x^\diamond \|^2+\| y_{T,k} -y^*(x^\diamond)  \|^2 ) \\
	& +3N^2\ell^2_{\theta,0} \| \hat{Z}_{B,k}- \boldsymbol{I}_m \otimes Z^*(x^\diamond) \|^2 \\
	&\leq \Xi_4 \|x_{k}-x^\diamond \|^2 +  \Xi_5 C_T \| y_{0,k}- y^*(x_k) \|^2 + 18mN^2\ell^2_{\theta,0}C_D\\
	& \| Z_{0,k}  - Z^R_k      \|^2_F  +18N^2\ell^2_{\theta,0}C_B \|\hat{Z}_{0,k}   - \boldsymbol{I}_m \otimes   Z_{D,k}   \|^2_F.
\end{split}
\end{equation}
\normalsize
The proof is finished. \hfill $ \blacksquare $
\subsection{Proof of \textbf{Proposition \ref{lemmaofx0z0zhat0}}}\label{proofoflemmaofx0z0zhat0}
	 Foremost, in light the warm start mechanism, i.e., $ y_{0,k} = y_{T,k-1} $,  we have  
	 \small
	 \begin{equation}
	 	\begin{split}
	 		& \| y_{0,k}- y^*(x_k) \|^2=  \| y_{T,k-1}- y^*(x_{k-1}) + y^*(x_{k-1}) - y^*(x_{k}) \|^2    \\   \notag
	 		&\leq 2C_T \| y_{0,k-1}- y^*(x_{k-1}) \|^2 + 2N\kappa^2 \|  x_{k} - x_{k-1}  \|^2 , 
	 	\end{split}
	 \end{equation}
 \normalsize
 Secondly, by $ Z_{0,k} =Z_{D,k-1} $, we have
 	 \small
 \begin{equation}
 	\begin{split}
 		 \| Z_{0,k}  - Z^R_k \|^2_F&=  \|  Z_{D,k-1} - Z^R_{k-1}  + Z^R_{k-1} - Z^R_k \|^2_F    \\   \notag
 		& \leq   2C_D \| Z_{0,k-1} - Z^R_{k-1} \|^2_F+ 2 \| Z^R_{k-1} - Z^R_k  \|^2_F . 
 	\end{split}
 \end{equation}
 \normalsize
 By substituting the definition of $ Z^R_k$, it yields 
  	 \small
 \begin{equation}
 	\begin{split}    
 		&  \| Z^R_{k-1} - Z^R_k  \|_F^2 = \| J_k H^{-1}_k - J_k H^{-1}_{k-1} + J_k H^{-1}_{k-1} - J_{k-1} H^{-1}_{k-1}  \|^2_F   \\   \notag
 		& \leq \frac{\Xi_1}{3m}(\|y_{T,k}-y_{T,k-1}\|^2+N\|x_k-x_{k-1}\|^2), 
 	\end{split}
 \end{equation}
 \normalsize
 and 
  \small
  \begin{equation}
 	\begin{split}    
 		&  \|y_{T,k}-y_{T,k-1}\|^2 = \|y_{T,k}- y^*(x_k) +y^*(x_k ) - y_{0,k}\|^2 \leq         \\ \notag
 		& (2+2C_T) \| y_{0,k}   -   y^*(x_k )    \|^2 \leq^{(a)} 4\| y_{0,k}   -   y^*(x_k )    \|^2,
 	\end{split}
 \end{equation}
 \normalsize
 where $ (a) $ holds since $ C_T < 1 $ always holds by a proper range of $ \alpha $. 
 Thus, we can have  
 	 \small
 	\begin{equation}
 		\begin{split}    
 			& \| Z_{0,k}  - Z^R_k \|^2_F \leq 2 C_D\|Z_{0, k-1}-Z_{k-1}^R\|_F^2+\Xi_7\|x_k-x_{k-1}\|^2  \\\notag
 			&+\frac{16 \Xi_1 C_T}{3m}\|y_{0, k-1}-y^*(x_{k-1})\|^2 .
 		\end{split}
 	\end{equation}
 	\normalsize 
Thirdly, by $ \hat{Z}_{0,k} = \hat{Z}_{B,k-1} $, we have 
 	 \small
\begin{equation}
	\begin{split}    
		& \| \hat{Z}_{0,k}  - \boldsymbol{I}_m \otimes Z_{D,k}   \|^2_F \leq \|  \hat{Z}_{D,k-1}  - \boldsymbol{I}_m\otimes Z_{D,k-1} +  \boldsymbol{I}_m\otimes Z_{D,k-1} \\ \notag
		&- \boldsymbol{I}_m \otimes Z_{D,k}     \|_F^2 \leq 2 C_B \|  \hat{Z}_{0,k-1}  - \boldsymbol{I}_m\otimes Z_{D,k-1} \|^2_F  + 2 m \| Z_{D,k-1} \\
		&-Z_{D,k}  \|^2_F, 
	\end{split}
\end{equation}
\normalsize
and by $ Z_{0,k} = Z_{D,k-1} $, it yields
\small
\begin{equation}
	\begin{split}    
     & \|Z_{D,k}  -  Z_{D,k-1}  \|^2_F  \leq  \| Z_{D,k} -  Z^R_k  +   Z^R_k -  Z_{0,k}  \|^2_F  	\leq (2C_D +2)  \\  \notag
	 & \|   Z_{0,k}  -   Z^R_k   \|^2 _F \leq^{(b)} 4\|   Z_{0,k}  -   Z^R_k   \|^2 _F,
	\end{split}
\end{equation}
\normalsize
where $ (b) $ holds since $ C_D <1 $ always holds by a proper range of $ \gamma $.
Thus, we have 
\small
\begin{equation}
	\begin{split}    
		& \| \hat{Z}_{0,k}  - \boldsymbol{I}_m \otimes Z_{D,k}   \|^2_F   \leq   2 C_B \|  \hat{Z}_{0,k-1}  - \boldsymbol{I}_m\otimes Z_{D,k-1} \|^2_F   \\  \notag
		& + 16mC_D \| Z_{0,k-1} - Z^R_{k-1} \|^2_F + \frac{128\Xi_1 C_T}{3} \| y_{0,k-1} - y^*(x_{k-1}) \|^2 \\
		& + 8m \Xi_7 \| x_k - x_{k-1} \|^2
	\end{split}
\end{equation}
\normalsize
Moreover, 
\small
\begin{equation} 
	\begin{split}    
		& \|x_k-x_{k-1}\|^2  = \|  \Pi_{\Omega} ( x_{k-1} - \beta_k \hat{\Psi}(x_{k-1}) ) - x_{k-1}   \|^2\\
		&\leq \|  \beta_k \hat{\Psi}(x_{k-1})   \|^2 =  \beta_k^2 \| \hat{\Psi} (x_{k-1}) - \Psi(x_{k-1}) + \Psi(x_{k-1})   \|^2    \\ \notag
		&  \leq 2\beta_k^2 \| \hat{\Psi} (x_{k-1}) - \Psi(x_{k-1})\|^2 +2 \beta_k^2 \| \Psi(x_{k-1}) \|^2    \\ 
		&  \leq  2\beta_k^2 \Xi_5 C_T \| y_{0,k-1}- y^*(x_{k-1}) \|^2 + 36\beta_k^2mN^2\ell^2_{\theta,0}C_D \| Z_{0,k-1} \\
		&    - Z^R_{k-1}      \|^2_F+36\beta_k^2N^2\ell^2_{\theta,0}C_B \|\hat{Z}_{0,k-1}   - \boldsymbol{I}_m \otimes   Z_{D,k-1}  \|^2_F + 2\beta_k^2   \Xi_6.\\   
	\end{split}
\end{equation}
\normalsize
Furthermore, by substituting $  \|x_k-x_{k-1}\|^2  $,  we have 
\small
\begin{equation}
	\begin{split}    
       &  \| y_{0,k}- y^*(x_k) \|^2  +   \| Z_{0,k}  - Z^R_k \|^2_F + \| \hat{Z}_{0,k}  - \boldsymbol{I}_m \otimes Z_{D,k}   \|^2_F    \\ \notag
       & \leq (2+\frac{16\Xi_3}{m} + \frac{128\Xi_1}{3} +2 \beta^2_k \Xi_5 \Xi_8 )C_T \|   y_{0,k-1}- y^*(x_{k-1})   \|^2 \\
       & +(2+16m+36 m N^2\beta^2_k \ell^2_{\theta,0} \Xi_8 ) \| Z_{0,k-1}  - Z^R_{k-1} \|^2_F  \\
       & + (2  + 36 N^2\beta^2_k \ell^2_{\theta,0} \Xi_8  )\| \hat{Z}_{0,k-1}  - \boldsymbol{I}_m \otimes Z_{D,k-1}   \|^2_F + 2\beta_k^2 \Xi_6 \Xi_8.
	\end{split}
\end{equation}
\normalsize	 
The proof is finished. \hfill $ \blacksquare $
\subsection{Proof of \textbf{Theorem \ref{theorem1}}}\label{proofoftheorem1} 
	Foremost, we have
	\small
	\begin{equation}
		\begin{split}
			&\|x_{k+1}-x^\diamond \|^2  \leq \| \Pi_{\Omega} (x_{k}-\beta_k\hat{\Psi}(x_{k})  -  \Pi_{\Omega} ( x^\diamond - \beta_k \Psi(x^\diamond) ) \|^2 \\
			&\leq  \|x_{k}-x^\diamond -\beta_k( \hat{\Psi}(x_{k}) -  \Psi(x^\diamond))\|^2  \leq \|x_{k}-x^\diamond \|^2  \notag  \\
			&  -2 \beta_k \left\langle x_{k} - x^\diamond, \hat{\Psi}(x_{k}) -  \Psi(x^\diamond) \right\rangle  +\beta^2_k \|  \hat{\Psi}(x_{k}) -  \Psi(x^\diamond)  \|^2  
		\end{split}
	\end{equation}
\normalsize
For the second term of the above inequality, we have 
\small
\begin{equation}
	\begin{split}
	  &	-2 \beta_k \left\langle x_{k} - x^\diamond, \hat{\Psi}(x_{k}) -  \Psi(x^\diamond) \right\rangle  
	  = -2 \beta_k \left\langle x_{k} - x^\diamond, \right.  \\ \notag
	  &\left.\hat{\Psi}(x_{k}) -  \Psi(x_{k}) \right\rangle -2 \beta_k \left\langle x_{k} - x^\diamond, \Psi(x_{k}) -  \Psi(x^\diamond) \right\rangle.   
	\end{split}
\end{equation}
\normalsize
where 
in conjunction with  Lemma \ref{lemmaofdistanceofz} and Lemma \ref{lemmaofx0z0zhat0}, it yields 
\small
\begin{equation}
	\begin{split}
		& \|  \hat{\Psi}(x_{k}) -  \Psi(x_{k})  \|^2   \notag \\
		& \leq ( \Xi_5 C_T + 18mN^2\ell^2_{\theta,0}C_D + 18N^2\ell^2_{\theta,0}C_B) ( \| y_{0,k}- y^*(x_{k}) \|^2     \\
		&  +  \| Z_{0,k}  - Z^R_{k}      \|^2_F  + \|\hat{Z}_{0,k}  - \boldsymbol{I}_m \otimes   Z_{D,k}  \|^2_F)   \leq ( \Xi_5 C_T \\
		& + 18mN^2\ell^2_{\theta,0}C_D + 18N^2\ell^2_{\theta,0}C_B)(  (\frac{1}{\pi})^k \Delta_{0} + 2\beta_k^2 \sum_{s=0}^{k-1} (\frac{1}{\pi})^{s} \Xi_6\Xi_8),
	\end{split}
\end{equation}
\normalsize
which implies 
\small
\begin{equation}
	\begin{split}
  &\|  \hat{\Psi}(x_{k}) -  \Psi(x_{k})  \|  \leq  \sqrt{C\Delta_{0}}(\frac{1}{\pi})^{k/2}  + \beta_k \sqrt{2C\sum_{s=0}^{k-1} (\frac{1}{\pi})^{s} \Xi_6\Xi_8}\\
  &\leq   \sqrt{C\Delta_{0}}(\frac{1}{\pi})^{k/2}  + \beta_k \sqrt{\frac{2\pi C\Xi_6\Xi_8}{\pi-1}} + \beta_k\sqrt{\frac{2\pi C \Xi_6\Xi_8}{\pi-1}}(\frac{1}{\pi})^{k/2}.  \notag
	\end{split}
\end{equation}
\normalsize
Thus, we have 
\small
\begin{equation}
\begin{split}
	&-2 \beta_k \left\langle x_{k} - x^\diamond, \hat{\Psi}(x_{k}) -  \Psi(x_{k}) \right\rangle \leq (1 + \|x_{k} - x^\diamond\|^2 ) \notag \\
	&(  \sqrt{C\Delta_{0}} \beta_k(\frac{1}{\pi})^{k/2}  + \beta^2_k \sqrt{\frac{2\pi C\Xi_6\Xi_8}{\pi-1}} + \beta^2_k\sqrt{\frac{2\pi C \Xi_6\Xi_8}{\pi-1}}(\frac{1}{\pi})^{k/2}   ).
\end{split}
\end{equation}
\normalsize
Let $ \eta_{1,k}:=\sqrt{C\Delta_{0}} \beta_k(\frac{1}{\pi})^{k/2}  + \beta^2_k \sqrt{\frac{2\pi C\Xi_6\Xi_8}{\pi-1}} + \beta^2_k\sqrt{\frac{2\pi C \Xi_6\Xi_8}{\pi-1}}(\frac{1}{\pi})^{k/2}    $, 
$ \eta_{2,k}:= \beta^2_k C ( (\frac{1}{\pi})^k\Delta_{0} + \frac{2\beta^2_k \Xi_6\Xi_8\pi}{\pi-1} + \frac{2\beta^2_k \Xi_6\Xi_8\pi}{\pi-1} (\frac{1}{\pi})^{k}   )    $, 
it follows that 
$ \sum_{k=0}^{\infty} \eta_{1,k}  < \infty   $,
$ \sum_{k=0}^{\infty} \eta_{2,k}  < \infty   $ .
Furthermore, in light of Assumption \ref{assumptionofstrictmonotone}, we have 
\small
\begin{flalign}
	\eta_{3,k} := -2 \beta_k \left\langle x_{k} - x^\diamond, \Psi(x_{k}) -  \Psi(x^\diamond) \right\rangle <0.
\end{flalign}
\normalsize
Thus, it follows that
\small
\begin{equation}
\begin{split}
 & \|x_{k+1}-x^\diamond \|^2 \leq  (1  + \eta_{1,k} ) \|  x_{k}  -x^{\diamond} \|^2_k + \eta_{2,k} +\eta_{1,k} -\eta_{3,k}.
\end{split}
\end{equation} 
\normalsize
where $ \sum_{k=0}^{\infty}\eta_{1,k} < \infty $ and  $ \sum_{k=0}^{\infty} (\eta_{1,k}+\eta_{2,k})  < \infty $.
In light of Lemma 5.31 in \cite{bauschke2011convex}, it can be verified that $ x_k  $ converges to $ x^\diamond $.
The proof is finished. \hfill $ \blacksquare $

\subsection{Proof of \textbf{Proposition \ref{proposition1}}}\label{proof_ofproposition1} 
	Foremost, in light of Lemma \ref{lemma_of_lipsmono},  we have 
	\small
	\begin{equation}
	    \begin{split}
	    	& \|  x_{k+1}  -  x^\diamond\| = \| \Pi_{\Omega} (x_{k}-\beta\hat{\Psi}(x_{k})  -  \Pi_{\Omega} ( x^\diamond - \beta \Psi(x^\diamond) ) \| \\ \notag 
	    	& \leq \|   x_{k} -x^\diamond -\beta (  \Psi(x_{k})  -  \Psi(x^\diamond)   )    \|   + \beta \|  \Psi(x_{k}) - \hat{\Psi} (x_{k})   \| \\
	    	&\leq G(\beta) \| x_{k} -  x^\diamond \| +\beta \tilde{C}_T \| y_{0,k} - y^*(x_{k}) \|\\
	    	&+ \beta \tilde{C_D} \| Z_{0,k}  - Z^R_{k} \|_F  + \beta \tilde{C_B}  \|\hat{Z}_{0,k}  - \boldsymbol{I}_m \otimes   Z_{D,k}  \|_F,  \\
	    \end{split}
	\end{equation}
\normalsize
where $ G(\beta) $, $ \tilde{C}_T $, $ \tilde{C}_D $, and $ \tilde{C}_B $ are defined in (\ref{notation2}).
In light of Lemma \ref{lemmaofoptimalcondition},  we have $ \Psi (x^\diamond) =0  $ since $ \Omega=\mathbb{R}^q $. 
Furthermore,  in conjunction with Lemma \ref{lemmaofdistanceofz}, 
the conclusion can be deduced. \hfill $ \blacksquare $
\subsection{Proof of \textbf{Theorem \ref{theorem2}}}\label{proof_oftheorem2} 
 \par Let $\beta=0$, we have a matrix $\boldsymbol{M}(0) $ with $w=[1 \quad 0\quad 0\quad 0]^\top$ and $v=[1\quad 0\quad 0\quad 0]^\top$ being 
 its left and right eigenvector respectively. By Theorem 6.3.12 in \cite{horn2012matrix}, 
 $\frac{d\lambda(\beta)}{d\beta}\left|_{\beta=0} \right. < 0$, which implies that 
 there must exists a positive $\beta$ close to $0$ such that $\rho(\boldsymbol{M}(\beta))<1$ holds. 
 Furthermore, $ \text{det} (  I_4   -\boldsymbol{M}(\beta) ) =0 $ indicates that  $\rho(\boldsymbol{M}(\beta))=1$.
Suppose $\beta_s$ is the smallest positive solution,  then $ \text{det} (  I_4   -\boldsymbol{M}(\beta_s) ) =0 $ holds. Thus, let $   0< \beta < \text{min}\{ \beta_{s} , \frac{2 m_{\theta}}{\ell_{\Phi}^2} \} $ holds, and the iteration counts $ T $, $ D $, and $ B $ satisfy the  (\ref{iterationcounts}) in Lemma \ref{lemmaofx0z0zhat0}, then $\rho(\boldsymbol{M}(\beta))<1$ holds. 
 Furthermore, in conjunction with the strong monotonicity of pseudo-gradient further indicates  $ \{ x_k \}_0^\infty  $ converges to a Nash equilibrium $x^\diamond$ of leaders, and $(x^\diamond, y^*(x^\diamond)) $ is a SE of (\ref{gamemodel}). The proof is finished. \hfill $ \blacksquare $

\subsection{Proof of \textbf{Lemma  \ref{additional_dual}}} \label{proof_ofadditional_dual}
\par By the strong convexity of $s^i(y^i,x)$, we have 
$s^i(y^{i,*}_{\vartheta^i}(x),x)  \leq  s^i(y^{i,*}(x),x) + \nabla_{y^i} s^i(y^{i,*}(x),x) ^{\top} ((y^{i,*}_{\vartheta^i}(x)  -  y^{i,*}(x) )  +  \frac{\mu}{2} \| (y^{i,*}_{\vartheta^i}(x)  -  y^{i,*}(x) \|^2$. 
Invoking by the approach of interior-point introduced in \cite{boyd2004convex}, we have  $ s^i(y^{i,*}_{\vartheta^i}(x),x) - s^i(y^{i,*}(x),x) \leq \frac{\tilde{s}}{\vartheta^i} $. And in conjunction with $ \nabla_{y^i} s^i(y^{i,*}(x),x) =\boldsymbol{0}$, it follows that $\|y^{i,*}_{\vartheta^i}(x)  -  y^{i,*}(x) \| \leq \sqrt{\frac{2\tilde{s}}{\mu \vartheta^i}}$.
Notably, it can be verified that $\|  \theta^j(x, y^*(x)) - \theta^j(x, y^*_\vartheta (x) )  \| \leq \ell_{\theta,0} \sqrt{\sum_{i=1}^{N} \frac{2\tilde{s}}{\mu \vartheta^i}} $.  The proof is complete. \hfill $ \blacksquare $

\subsection{Proof of \textbf{Proposition  \ref{implict_gradient}}} \label{proof_ofimplict_gradient}
\par 
The Lagrangian function of  (\ref{newfollowerproblem})
is given by $L^i(y^i,x,\lambda)=s_R^i(y^i,x)+\lambda^\top(A^iy^i-b^i)$, and it can be inferred that 
$\nabla_{y^{i}}L^{i}(x,y^{i,\diamond},\lambda^{\diamond})=\nabla_{y^{i}}s_{R}^{i}(y^{i,\diamond},x)+\left(A^{i}\right)^{\top}\lambda^{\diamond} = \boldsymbol{0}$ and 
$\nabla_\lambda L^i(x,y^{i,\diamond},\lambda^{\diamond})=A^iy^{i,\diamond}-b^i=\boldsymbol{0}$. 
In conjunction with Lemma 4.2 in \cite{gould2016differentiating}, it can be further 
deduced that $\nabla_xy^{i,*}(x)\nabla_{y^i}^2s_R^i(y^{i,*}(x),x)+\nabla_x\nabla_{y^i}s_R^i(y^{i,*}(x),x)+\nabla_x\lambda^*(x)A^i=\boldsymbol{0}$ and $\nabla_xy^{i,\diamond}(x)\left(A^i\right)^{\top}=\boldsymbol{0}$. Thus, it follows that 
\small
\begin{equation}\label{lasteq}
	\begin{aligned}
		\begin{bmatrix}\nabla_{y^i}^2s_R^i{\left(y^{i,*}(x),x\right)}&{\left(A^i\right)}^\top\\A^i&\boldsymbol{0}
		\end{bmatrix}
		\begin{bmatrix}\nabla_xy^{i,*}(x)^\top\\\nabla_x\lambda^*(x)
		\end{bmatrix}\\
		+\begin{bmatrix}
			\nabla_x\nabla_{y^i}s_R^i{\left(y^{i,*}(x),x\right)}^\top\\\boldsymbol{0}
		\end{bmatrix}
		=\begin{bmatrix}\boldsymbol{0}\\\boldsymbol{0}\end{bmatrix}.\\
	\end{aligned}
\end{equation}
\normalsize
By solving (\ref{lasteq}), we have 
$\nabla_x\lambda^*(x)=-\left[A^i{\left(H_R^{i,*}\right)}^{-1}{\left(A^i\right)}^{\top}\right]^{-1}A^i{\left(H_R^{i,*}\right)}^{-1}\nabla_x\nabla_{y^i}s_R^i{\left(y^{i,*}(x),x\right)}^{\top}$ and 
\small
\begin{equation}
	\begin{aligned}
		&\nabla_xy^{i,s}(x)^{\top}=\left(-\left(H_R^{i,*}\right)^{-1}+\left(H_R^{i,*}\right)^{-1}\left(A^i\right)^{\top}\right.\\
		&\left.\left[A^i\left(H_R^{i,*}\right)^{-1}\left(A^i\right)^{\top}\right]^{-1}A^i\left(H_R^{i,*}\right)^{-1} \right) \nabla_x\nabla_{y^i}s_R^i\left(y^{i,*}(x),x\right)^{\top}. \notag
	\end{aligned}
\end{equation} 
\normalsize
Thus, the proof is finished. \hfill $ \blacksquare $

%

\ifCLASSOPTIONcaptionsoff
  \newpage
\fi

\bibliographystyle{IEEEtran}
\bibliography{IEEENEW}

\end{document}